\documentclass[useAMS,usenatbib]{mn2e}
\usepackage{graphicx}
\usepackage{subfigure}

\title[Superluminal models of the spectra of 9 broad-band 
pulsars]
{Comparison of multiwavelength observations of 9 broad-band 
pulsars with the spectrum of the emission from an extended
current with a superluminally rotating distribution pattern}
\author[H. Ardavan et al.]{H.\ Ardavan,$^1$
A.\ Ardavan,$^2$ J.\ Singleton,$^3$ J.\ Fasel,$^4$ W.\ Junor,$^5$
\newauthor J.\ Middleditch,$^6$ M.\ R.\ Perez,$^7$ A.\ Schmidt,$^4$ 
P. Sengupta,$^{3,8}$ P.\ Volegov$^9$\\
$^1$Institute of Astronomy, University of Cambridge,
Madingley Road, Cambridge CB3 0HA, UK\\
$^2$Clarendon Laboratory, Department of Physics, University of Oxford,
Parks Road, Oxford OX1 3PU, UK\\
$^3$MPA-NHMFL, MS-E536,
Los Alamos National Laboratory, Los Alamos, New Mexico 87545, USA\\
$^4$AET-2, MS-E548,
Los Alamos National Laboratory, Los Alamos, New Mexico 87545, USA\\
$^5$ISR-2, MS-D436,
Los Alamos National Laboratory, Los Alamos, New Mexico 87545, USA\\
$^6$CCS-3, MS-B265,
Los Alamos National Laboratory, Los Alamos, New Mexico 87545, USA\\
$^7$Astrophysics Division, 3Y28, NASA Headquarters,
330 E.\ Street SW, Washington DC 20546, USA\\
$^8$T-4, T-CLNS, MS-B258,
Los Alamos National Laboratory, Los Alamos, New Mexico 87545, USA\\
$^9$P-21, MS-D454,
Los Alamos National Laboratory, Los Alamos, New Mexico 87545, USA}

\begin{document}

\date{June 2009}

\pagerange{\pageref{firstpage}--\pageref{lastpage}} \pubyear{2009}

\maketitle

\label{firstpage}

\begin{abstract}
The observed spectra of 9 pulsars for which 
multiwavelength data are available from radio to 
$X$- or $\gamma$-ray bands
(Crab, Vela, Geminga, B0656+14, B1055-52, 
B1509-58, B1706-44, B1929+10, and B1951+32) 
are compared with the spectrum of 
the radiation generated by an extended polarization current whose 
distribution pattern rotates faster than light {\it in vacuo}.
It is shown that by inferring the values of two free parameters from 
observational data (values that are consistent with those of 
plasma frequency and electron cyclotron frequency in a conventional 
pulsar magnetosphere), and by adjusting the spectral indices of 
the power laws describing the source spectrum in various frequency 
bands, one can account {\em quantitatively} for the entire spectrum 
of each pulsar in terms of a single emission process. 
This emission process (a generalization of the synchrotron-\'Cerenkov 
process to a volume-distributed source in vacuum) gives rise to
an oscillatory radiation spectrum. Thus,
the bell-shaped peaks of pulsar spectra in the ultraviolet 
or $X$-ray bands (the features that are normally interpreted  
as manifestations of thermal radiation) appear in the present 
model as higher-frequency maxima of the same oscillations that
constitute the emission bands observed in the radio spectrum of the 
Crab pulsar.  Likewise, the sudden steepening of the gradient
of the spectrum by $-1$, which occurs 
around $10^{18}-10^{21}$ Hz, appears as a universal feature of 
the pulsar emission: a feature that reflects 
the transit of the position of the observer across the 
frequency-dependent Rayleigh distance.   
Inferred values of the free parameters of the present model 
suggest, moreover, that the lower the rotation 
frequency of a pulsar, the more weighted towards higher 
frequencies will be its observed spectral intensity.  

\end{abstract}

    
\section{Introduction}
\label{sec:1}

Pulsars are unique in emitting a radiation whose spectrum 
extends from radio waves to gamma rays with essentially 
the same characteristics~\citep{b17}. The fact that in many 
pulsars the pulses emitted over widely separated frequency bands
(e.g.\ in the radio and $X$-ray bands) are closely aligned in phase  
and have correlated profiles~\citep{b28,b29,b30,b35,b31}  
indicates that both 
the mechanism by which such pulses are generated and the 
magnetospheric site from which they originate must be the same 
for all frequencies.  This notwithstanding, the extant 
models of pulsar radiation invoke a diverse set of unrelated 
emission mechanisms and emission sites to account for the 
observational data on multiwavelength characteristics of 
this radiation~\citep{b32,b36,b33,b34}.

It was shown in~\citet{b1} that the rigid rotation of the overall distribution 
pattern of the pulsar emission can only arise from an emitting current
whose distribution pattern likewise rotates rigidly with the same 
angular frequency (see also Section 2).  The fact that this requires
the pattern of distribution of the emittting current to move with a 
linear speed exceeding the speed of light {\it in vacuo} outside 
the pulsar's light cylinder is not 
incompatible with special relativity: the charge separation resulting 
from the coordinated motion of an aggregate of charged particles gives 
rise to a polarization current whose distribution pattern can have a 
superluminal motion~\citep{b3,b4,b14}. Moreover, Maxwell's 
equations show that a polarization current contributes towards
the radiation field in just the same way as a current of free charges.     
In fact, such superluminal sources of radiation have been 
experimentally realized by several groups~\citep{b5,b7,b6,b8}. 

Here we show that the spectrum of the emission from a
polarization current whose distribution pattern rotates faster
than light {\it in vacuo} fits the spectra of 9 pulsars, 
for which multiwavelength data are available, over the 
entire range of their output from radio waves to gamma rays. 
The model we adopt for the source is a generic one: it has 
an azimuthally-fluctuating  
distribution pattern that both rotates and oscillates [as realized in 
recent experiments~\citep{b7,b6}], and it has sharp spatial fluctuations 
in the direction parallel to the rotation axis
with power-law spectra in various frequency bands
[as demanded by recent numerical models of the 
pulsar magnetosphere that predict the formation of 
current sheets~\citep{b2}].
We infer the values of the two adjustable parameters, 
characterizing the frequencies of temporal and azimuthal fluctuations
of the source, from the observational data, and choose 
the spectral indices of the power laws describing the spectral 
distribution of the variations of the source parallel to the 
rotation axis by optimizing the goodness of fit to the data.
We shall see that the theoretical spectra thus derived from a single 
emission process (a process that takes place in a localized region 
of the magnetosphere just outside its light cylinder) can
{\em quantitatively} account for the data on all nine of these 
extensively observed pulsars over 15 to 18 orders of magnitude of frequency.

The emission mechanism in question may be regarded as a 
generalization of synchrotron-\'Cerenkov process to a 
volume-distributed source {\it in vacuo}.  Its spectral 
distribution is described by the square of a Bessel function 
whose argument exceeds its order, and so has an oscillating 
amplitude with an algebraic (rather than exponential) 
rate of decay with frequency~\citep{b10}. 
Giving rise to a radiation whose spectrum is oscillatory, it is a 
mechanism that naturally accounts for~\citep{b1} the occurrence of the 
observed emission bands in the dynamic spectrum of the Crab 
pulsar~\citep{b16}.  

The unified explanation this mechanism
offers for the apparently unrelated features of pulsar spectra 
in different spectral bands 
sharply contrasts with the variety of models, proposed 
in the published literature, which could at best explain each 
of these features separately~\citep{b32,b36,b33,b34}.
The bell-shaped peak of the spectrum in the ultraviolet or $X$-ray bands 
that is normally attributed to black-body radiation from hot spots 
in the pulsar magnetosphere~\citep{b36}, for instance, emerges as
a higher-frequency maximum of the same  
proportionately-spaced oscillations of the spectrum 
that~\citet{b16} have observed in the radio band. 
Moreover, the decrease 
in the spectral index of the radiation by $-1$ that is universally 
encountered in pulsar spectra around $10^{18}-10^{21}$ Hz 
turns out to arise, not from any changes in the spectral distribution 
of the source of the radiation, but from a transition though
the Rayleigh distance (see Section~\ref{sec:4}).  
   
This paper is organized as follows.  In Section~\ref{sec:2.1}, we 
point out why the observational data on pulsars demand that the 
source of pulsar radiation should have a superluminally rotating 
distribution pattern. In Section 2.2, we describe the field due to
a constituent volume element of such a source, i.e.\ the
Li\'enard-Wiechert field generated by the 
superluminal counterpart of synchrotron process~\citep{b13,b11}.  
In Section~\ref{sec:2.3}, we consider the field arising from the 
entire volume of the source. 
We point out how the most efficient parts of a pulsar 
magnetosphere for the formation of signals
detectable at large distances
are the thin filaments within the superluminally rotating 
part of its current distribution pattern that approach the observer
with the speed of light and zero acceleration at the retarded
time~\citep{b12}.  Because the wave fronts 
from each such filament form a caustic on which the 
contributions that are made towards the field over a finite 
interval of emission time are received during a considerably 
shorter interval of observation time, we will see that the 
overall radiation beam generated by a volume source 
consists of an incoherent superposition of coherent, narrowing
subbeams whose intensities diminish as ${R_P}^{-1}$, instead of 
${R_P}^{-2}$, with the distance $R_P$ from their source~\citep{b11}.
Consequently, the radiation detected at large distances
from the pulsar will always be dominated by these subbeams.  
In Section \ref{sec:4}, we fit the 
multiwavelength observational data on each of the 9 pulsars
listed above with the frequency spectrum predicted for a rotating superluminal
source; our earlier derivation of the spectrum~\citep{b10} 
is outlined from an alternative point of view in Appendix A. 
Implications of the inferred values of the parameters of the  
fitted spectra will be discussed in Section~\ref{sec:5}, and a 
summary will be given in Section~\ref{sec:6}.

\section{Superluminal model of pulsars}
\label{sec:2}
\subsection{Observational constraints on the
motion of the distribution pattern of the source}
\label{sec:2.1}
The rigid rotation of the overall distribution pattern of the pulsar emission 
reflects a radiation field ${\bf E}$ whose cylindrical components depend 
on the cylindrical coordinates $(r,\varphi,z)$ and time $t$ as
\begin{equation}
E_{r,\varphi,z}(r,\varphi,z;t)=E_{r,\varphi,z}(r,\varphi-\omega t,z,t),
\label{eq:1}
\end{equation}
where $\omega$ is the rotation frequency of the pulsar.  Such a field can only 
arise from an electric current whose density ${\bf j}$ likewise depends on the 
azimuthal angle $\varphi$ in the combination $\varphi-\omega t$ only:
\begin{equation}
j_{r,\varphi,z}(r,\varphi,z;t)=j_{r,\varphi,z}(r,\varphi-\omega t,z,t)
\label{eq:2}
\end{equation}
[see Appendixes A and B of~\citet{b1}].  This property of the emitting current 
follows not only from the observational data, but also from the numerical models 
of the magnetospheric structure of an oblique rotator; it is found 
that any time-dependent structures in such models rapidly approach a steady state 
in the corotating frame~\citep{b2}. 

Unless there is no plasma outside the light cylinder, therefore, 
the emitting polarization current distribution in the magnetosphere of a pulsar should have a 
superluminally rotating pattern for $r>c/\omega$ (where $r$ is the radial distance 
from the axis of rotation and $c$ is the speed of light {\it in vacuo}).  
Such a source is not inconsistent with special relativity, as the
charged particles that make up the polarization current distribution
need only move relatively slowly~\citep{b3,b4}.  
It has been experimentally verified 
that such moving charged patterns 
act as sources of radiation in precisely the same way 
as any other moving sources of electromagnetic fields~\citep{b5,b6,b7,b8}.

Many distinctive features of the emission from a superluminal source are illustrated by the 
radiation from the following generic polarization current whose distribution pattern 
rotates and oscillates at the same time: ${\bf j}=\partial{\bf P}/\partial t$ for which
\begin{eqnarray}
P_{r,\varphi,z}(r,\varphi,z,t)&=&
s_{r,\varphi,z}(r,z)\cos(m{\hat\varphi})\cos(\Omega t),\nonumber\\
&&\qquad\qquad\qquad\qquad-\pi<{\hat\varphi}\le\pi,
\label{eq:3}
\end{eqnarray}
and ${\hat\varphi}\equiv\varphi-\omega t$.
Here, $P_{r,\varphi,z}$ are the components of the polarization ${\bf P}$
in a cylindrical coordinate system based on the axis of rotation,
${\bf s}(r,z)$ is an arbitrary vector function with a finite 
support in $r>c/\omega$,
$m$ is a positive integer, and $\Omega$ is an angular 
frequency whose value differs from an integral multiple of 
the rotation frequency $\omega$
[for the significance of this incommensurablity 
requirement, see Appendix A and \citet{b10}].
For a fixed value of $t$,
the azimuthal dependence of the polarization (\ref{eq:3})
along each circle of radius $r$ within the source
is the same as that of a sinusoidal wave 
train with the wavelength $2\pi r/m$
whose $m$ cycles fit around the 
circumference of the circle smoothly.
As time elapses,
this wave train both propagates around each 
circle with the velocity $r\omega$
and oscillates in its amplitude with the frequency $\Omega$.
This is a generic source:
one can construct any distribution
with a uniformly rotating pattern,
$P_{r,\varphi,z}(r,{\hat\varphi},z)$,
by the superposition over $m$
of terms of the form $s_{r,\varphi,z}(r,z,m)\cos(m{\hat\varphi})$.
It also corresponds to laboratory-based sources
that have been used in experimental demonstrations
of some of the phenomena described below~\citep{b6,b7}. 

The results reported here are derived from the retarded 
solution of Maxwell's equations for the above current 
distribution [see~\citet{b1,b12,b11,b10,b13,b9}].

\subsection{The field generated by a single volume element of the source}
\label{sec:2.2}

\begin{figure}
\centering
\includegraphics[height=7cm]{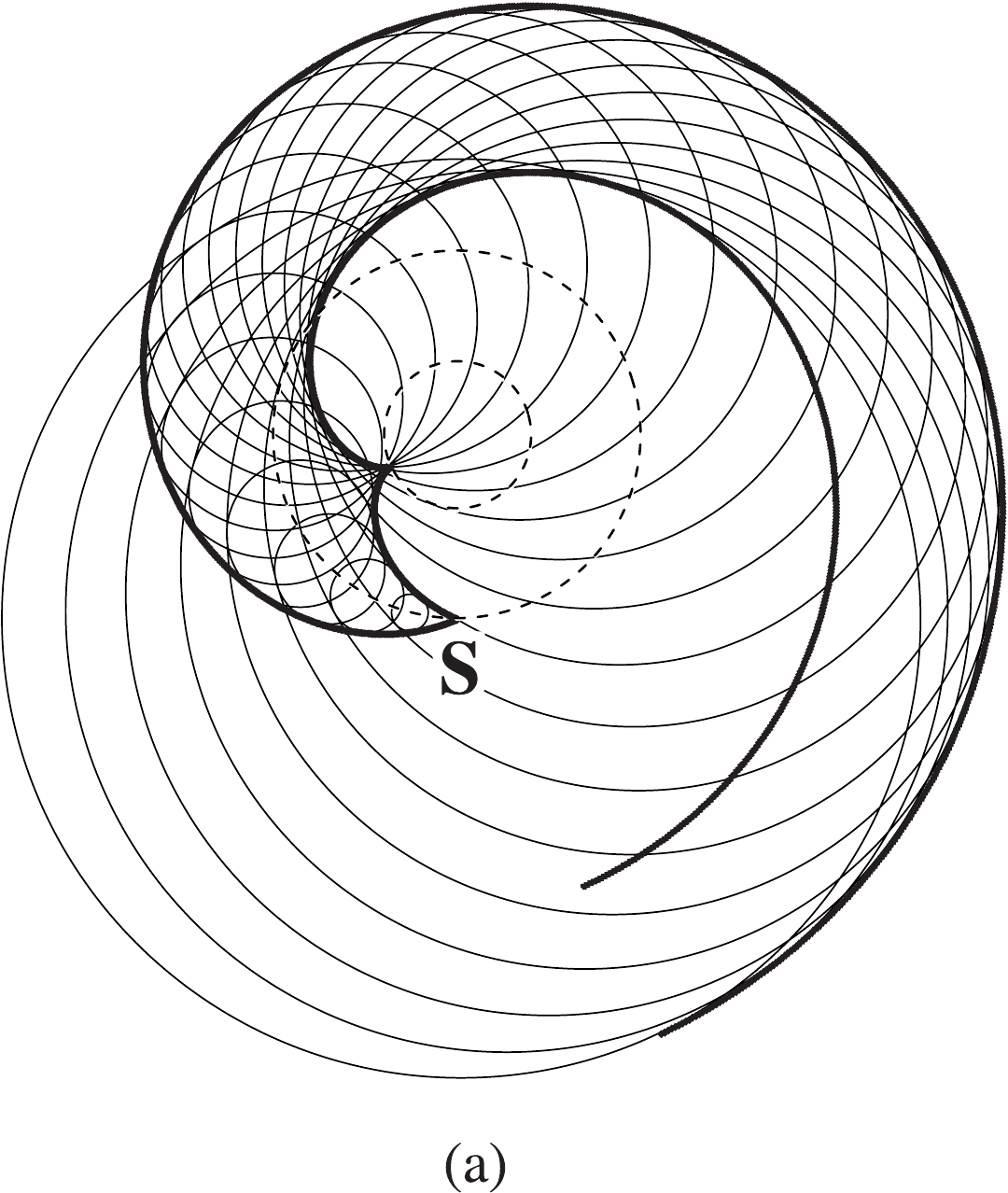} 
\subfigure{\includegraphics[height=3cm]{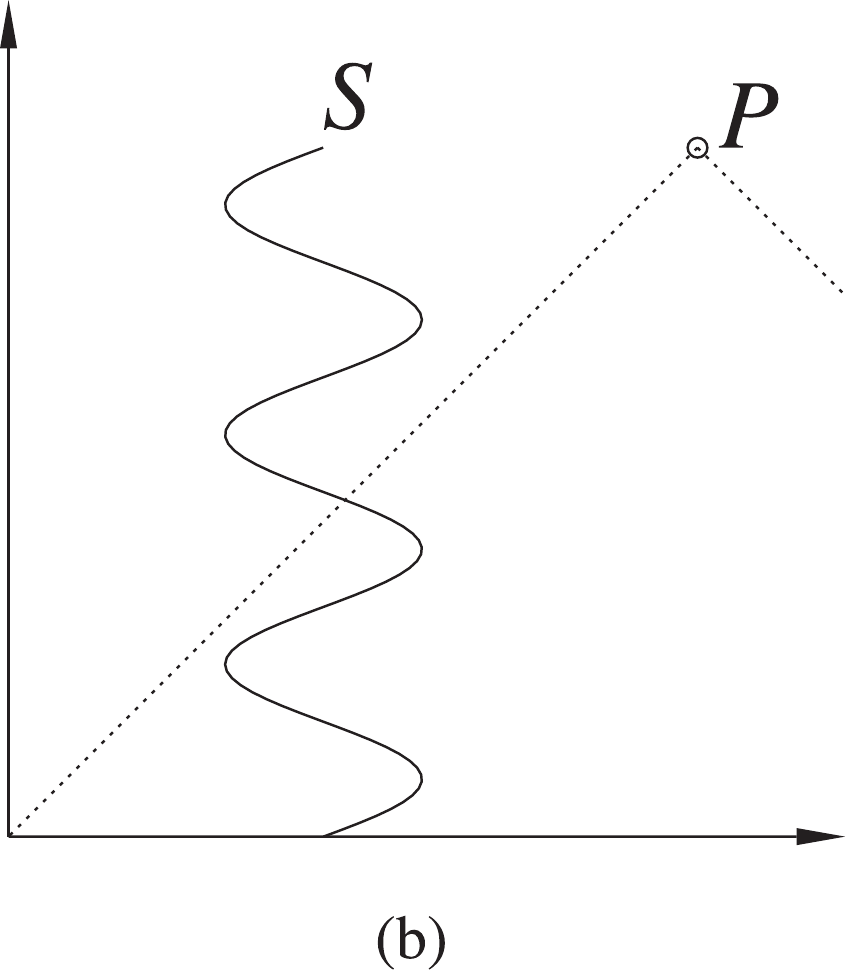}}
\subfigure{\includegraphics[height=3cm]{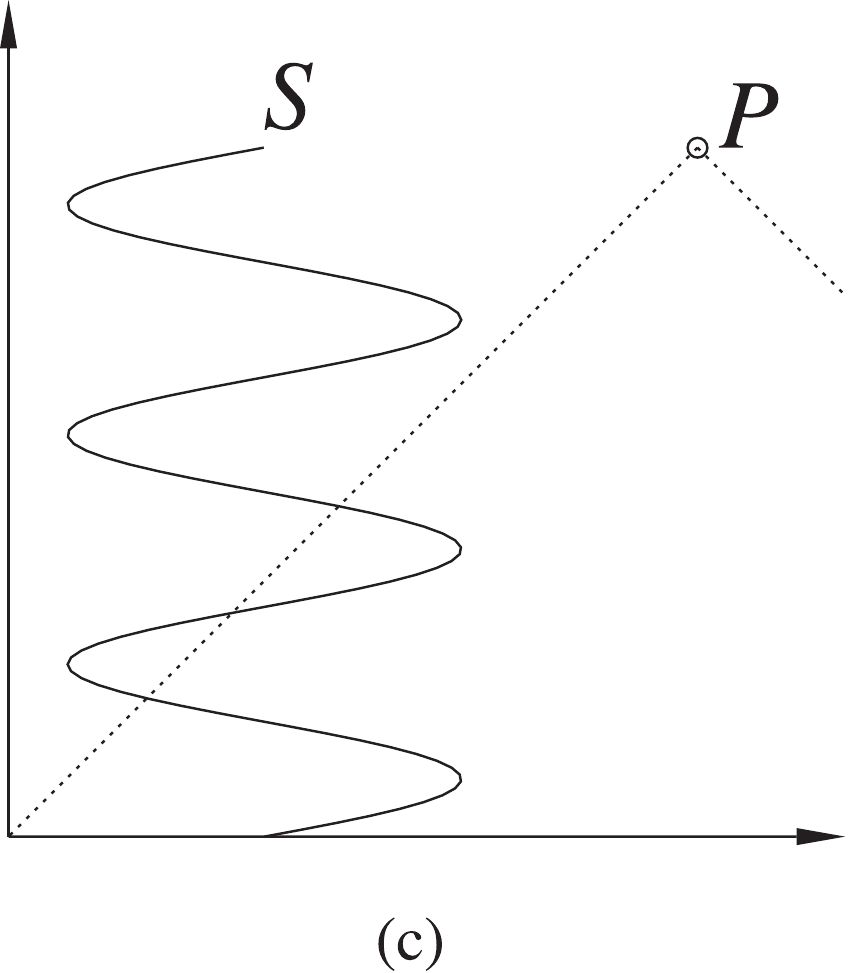}}
\subfigure{\includegraphics[height=3cm]{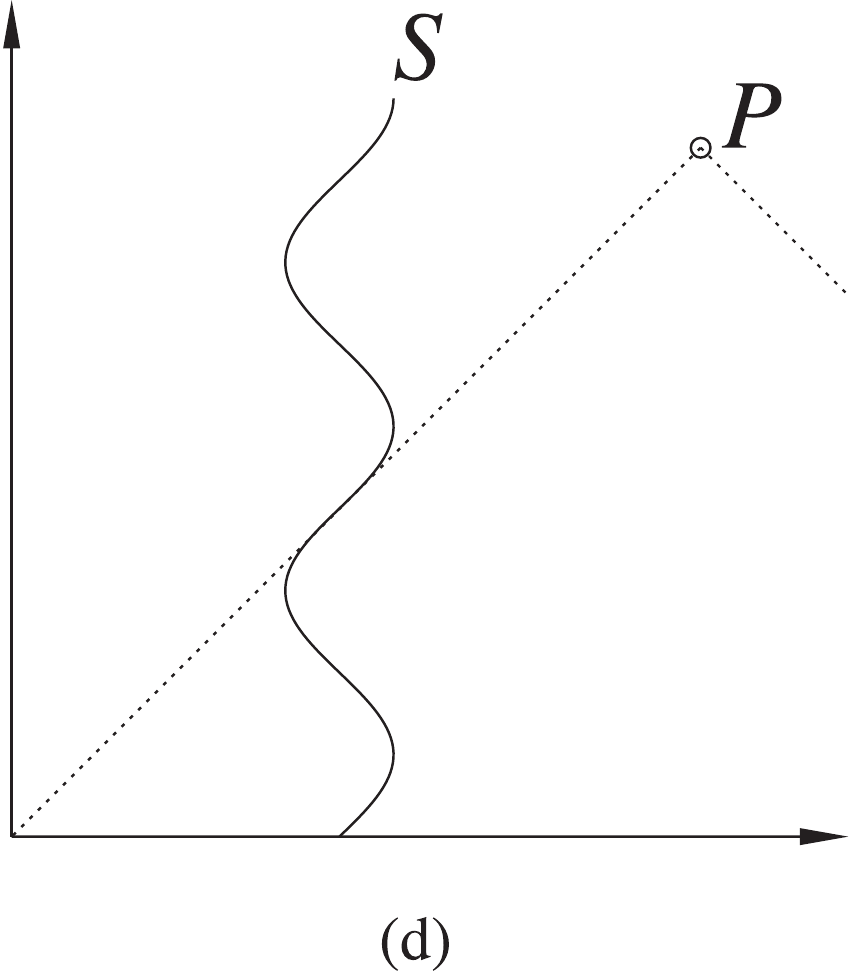}}
\caption{(a) Envelope of the spherical wave fronts 
emanating from a superluminally moving source 
element (S) in uniform circular motion.  The heavy curves 
show the cross section of the envelope with the 
plane of the orbit of the source.  The larger 
of the two dotted circles designates the orbit 
and the smaller the light cylinder.  (b), (c) and (d) Space-time 
($ct$ versus $x$) diagrams showing the intersection of the trajectory 
of the source point $S$ with the past light cone of the 
observation point $P$ when $P$ lies outside (b), inside (c), 
and on the cusp of (d) the envelope of wave fronts.}
\label{fig1}
\end{figure}

A superluminal source is necessarily volume-distributed~\citep{b4}.  
However, its field can be built up from the superposition of the fields 
of its moving constituent volume elements which are 
essentially point-like.  
Figure~\ref{fig1}(a) shows that the waves generated by a 
constituent volume element of a rotating superluminal source 
possess a cusped envelope and that, inside the envelope, 
{\it three} wave fronts pass through any given observation 
point simultaneously.  This reflects the fact that the field 
inside the envelope receives simultaneous 
contributions from three distinct values of the retarded time 
[see Fig.~\ref{fig1}(c)].  On the cusp of the envelope, 
where the space-time trajectory of the source is tangent 
to the past light cone of the observer [Fig.~\ref{fig1}(d)], 
all three contributions toward the value of the field 
coalesce~\citep{b9,b10,b11}.  

On this cusp (caustic), the source approaches the observer 
with the speed of light and zero acceleration at the retarded time, 
i.e.\ ${\rm d}R(t)/{\rm d}t=-c$ and ${\rm d}^2R(t)/{\rm d}t^2=0$, 
where $R(t)\equiv\vert{\bf x}(t)-{\bf x}_P\vert$ 
is the distance between the source point ${\bf x}(t)$ 
and the observation point ${\bf x}_P$. 
As a result, the interval of emission time 
for the signal carried by the cusp is 
much longer than the interval of its reception time~\citep{b10}.
This essentially instantaneous reception of contributions 
from an extended period of emission time represents focusing 
of the radiation in the time domain: a unique effect that has 
already been demonstrated experimentally~\citep{b6,b7}. 

\begin{figure} 
\centering
\includegraphics[height=4cm]{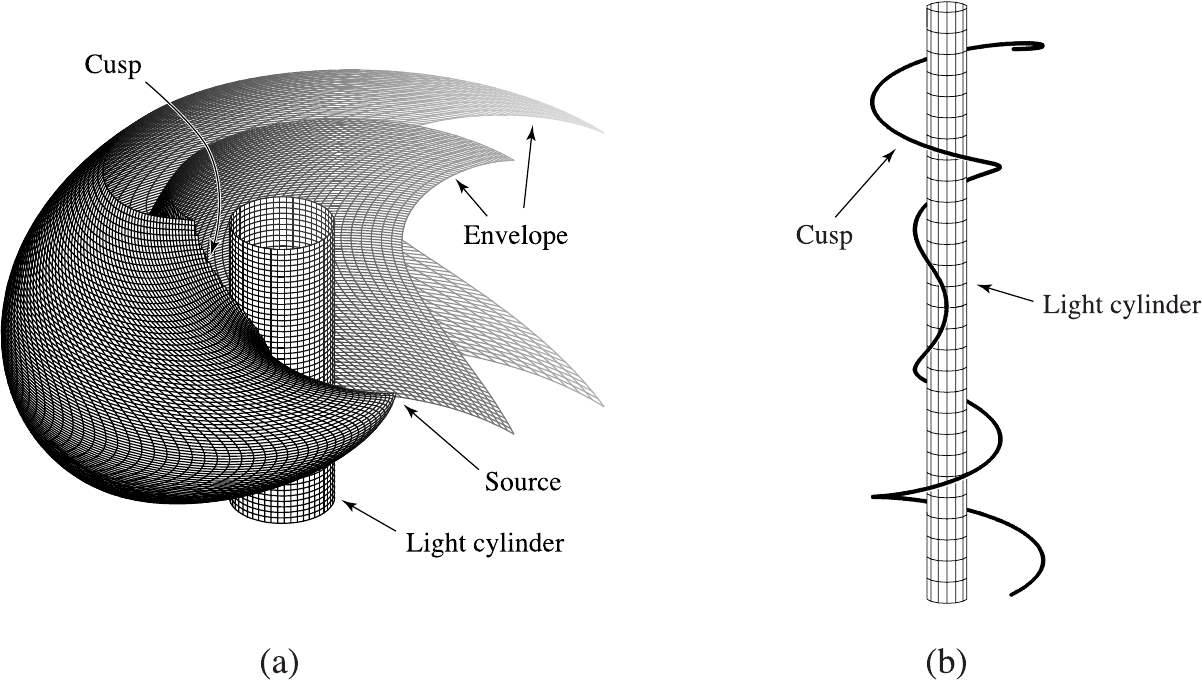}
\caption{Three dimensional views of the envelope (a) and its cusp (b).}
\label{fig2}
\end{figure}

A three-dimensional view of the envelope of wave fronts 
and its cusp is shown in Fig.~\ref{fig2}.  
The two sheets of the envelope, and the cusp along 
which these two sheets meet tangentially, 
spiral outward into the far zone.  
In the far zone, the cusp lies on the double cone 
$\theta_P=\arcsin[c/(r\omega)]$, $\theta_P=\pi-\arcsin[c/(r\omega)]$, 
where $(R_P, \theta_P, \varphi_P)$ denote the spherical 
polar coordinates of the observation point $P$.  
Thus, a stationary observer in the polar interval 
$\arcsin[c/(r\omega)]\le\theta_P\le\pi-\arcsin[c/(r\omega)]$ 
receives recurring pulses as the envelope rotates 
past him/her~\citep{b12}.

\begin{figure}
\centering
\includegraphics[height=7cm]{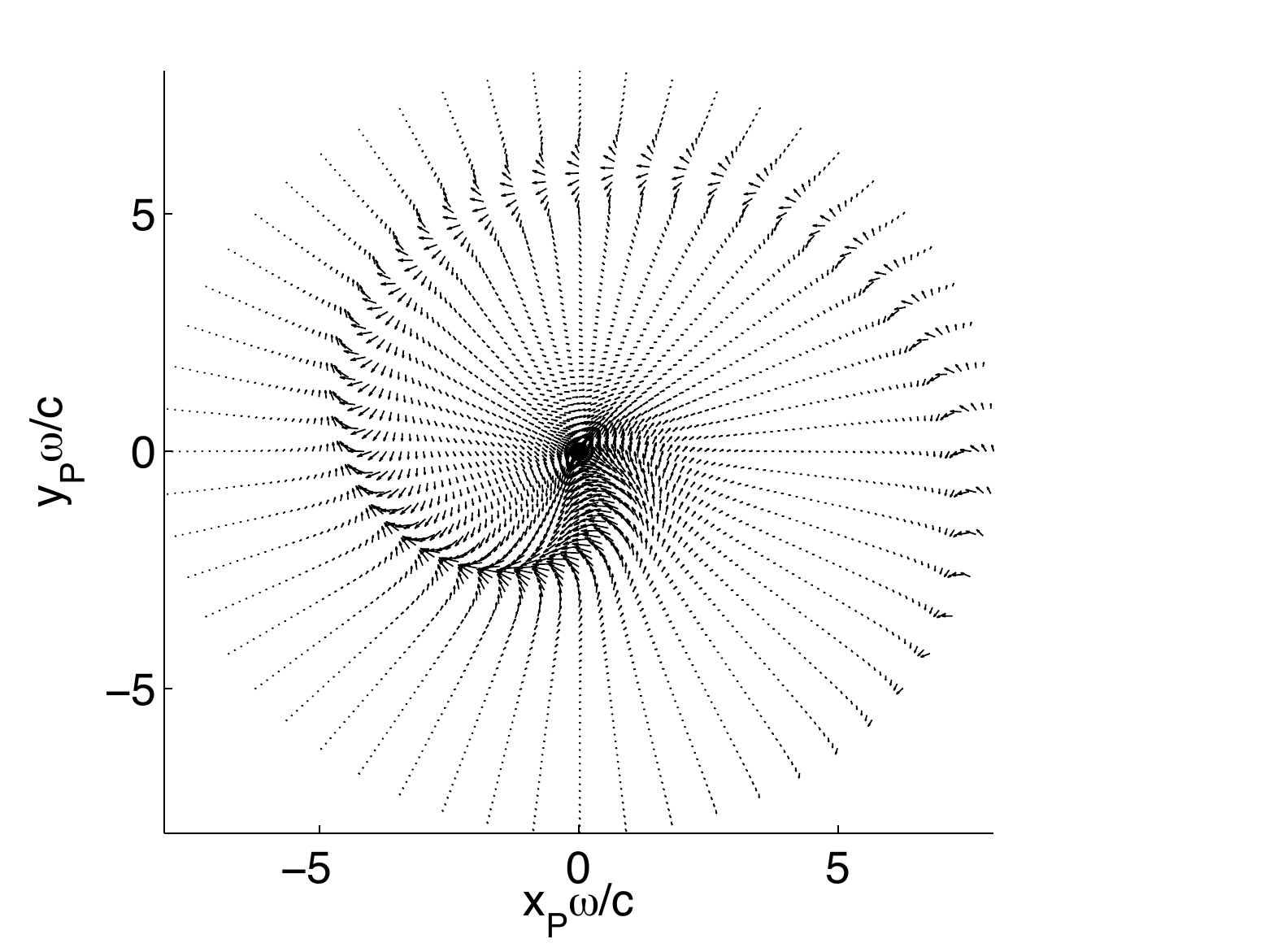}
\caption{Polarization position angles and field strengths 
on the cone $\theta_P=\pi/12$ outside the envelope 
of wave fronts for a source element with $r\omega=2$.
Here, the field vectors are projected onto the plane $(x_P,y_P)$
of the source's orbit.}
\label{fig3}
\end{figure}

Figure~\ref{fig3} shows the radiation field generated 
by the rotating source element $S$ on a cone close to the cusp, 
just outside the envelope.  Not only does the spiralling 
cusp embody a recurring pulse, but the plane of polarization 
of the radiation swings across the pulse~\citep{b13},
as in the radio emission received from pulsars~\citep{b17}.  

\begin{figure}
\centering
\includegraphics[height=5cm]{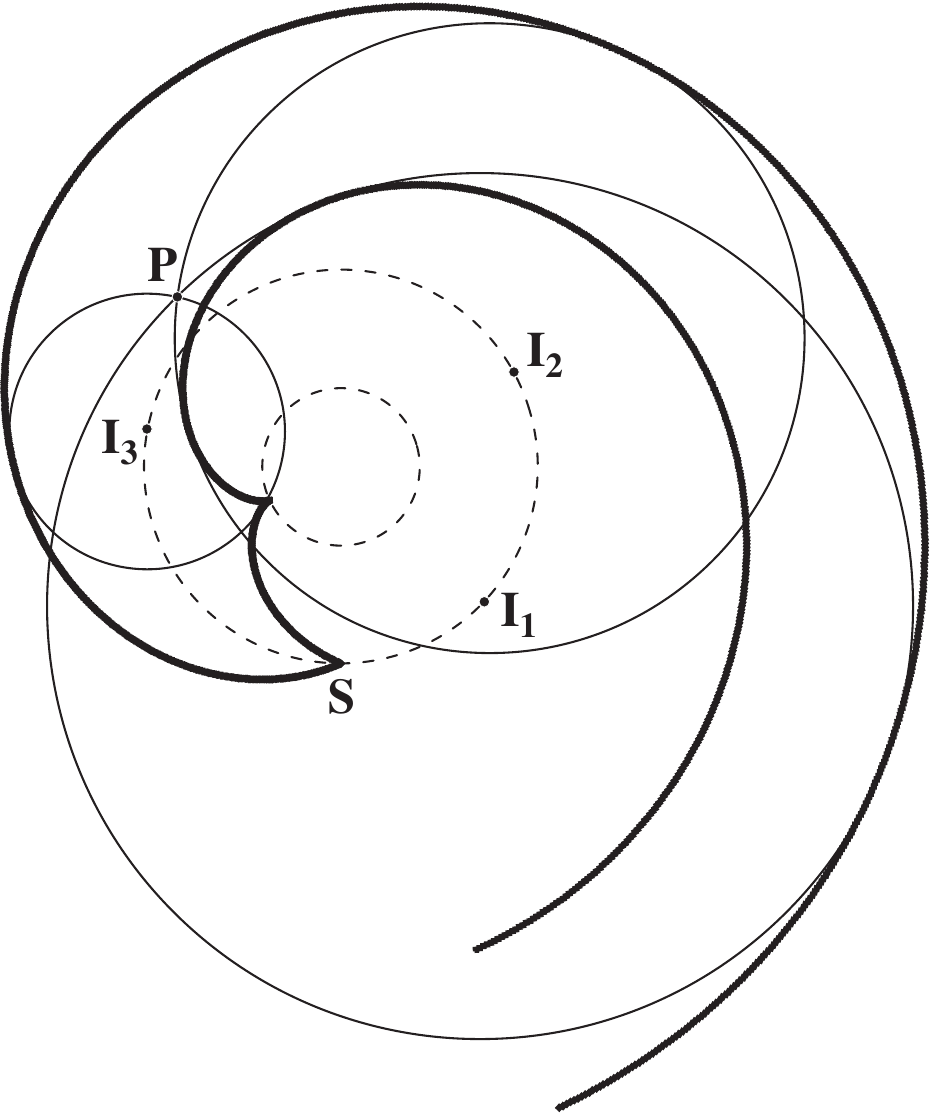}
\caption{An observer $P$ inside the envelope detects 
three images ($I_1$, $I_2$, $I_3$) of the source $S$ 
simultaneously.}
\label{fig4}
\end{figure}

As a consequence of the multivaluedness of the retarded time, 
three images of the source are observeable inside the envelope 
at any given observation time (Fig.~\ref{fig4}).  
The waves that were emitted when the source was at the retarded 
positions $I_1$, $I_2$ and $I_3$ in Fig.~\ref{fig4} 
are all received simultaneously at the observation point $P$.  
These images are detected as distinct components, or modes, 
of the emitted radiation~\citep{b14}. 

\begin{figure}
\centering
\subfigure{\includegraphics[width=7cm]{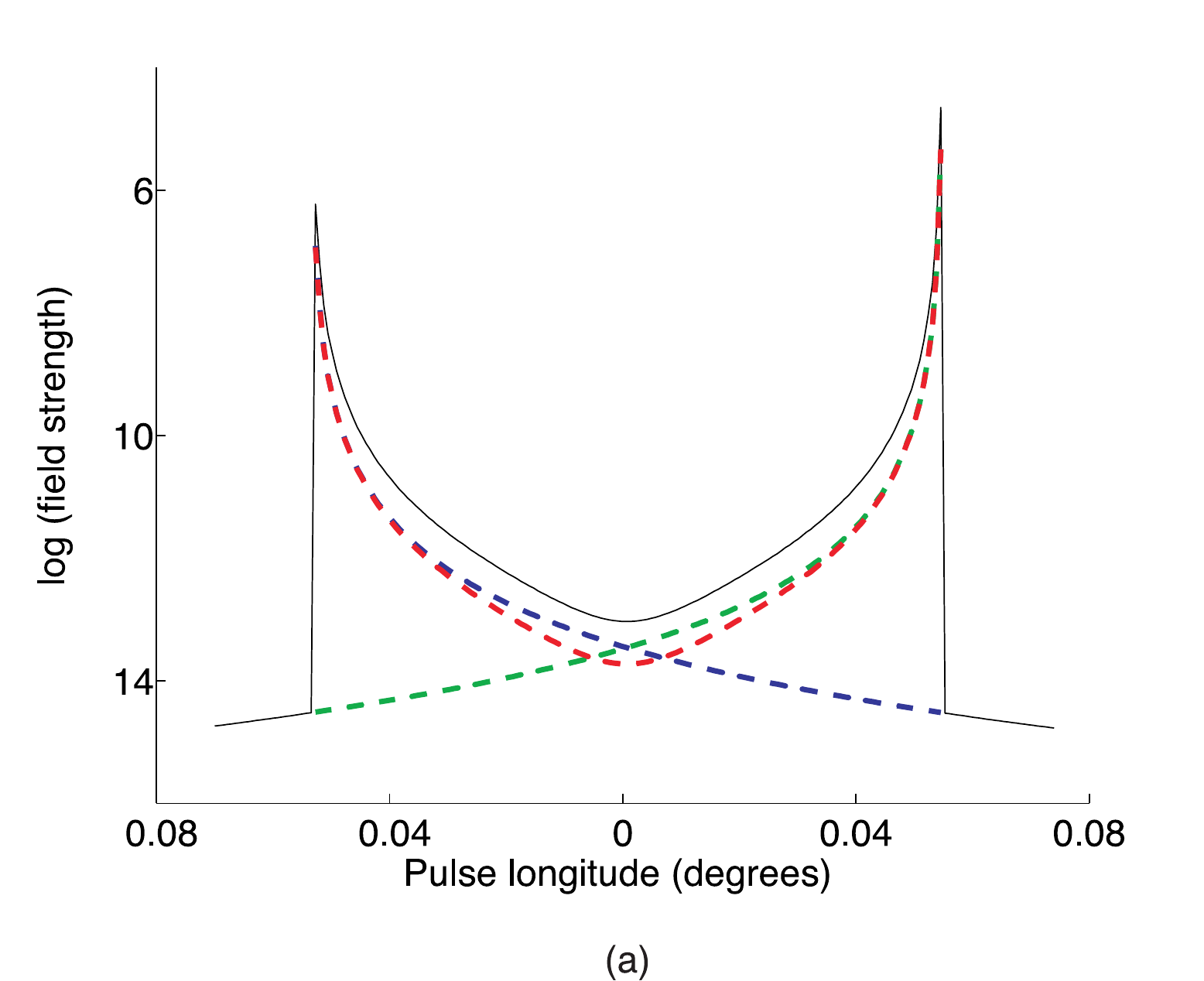}}
\subfigure{\includegraphics[width=7cm]{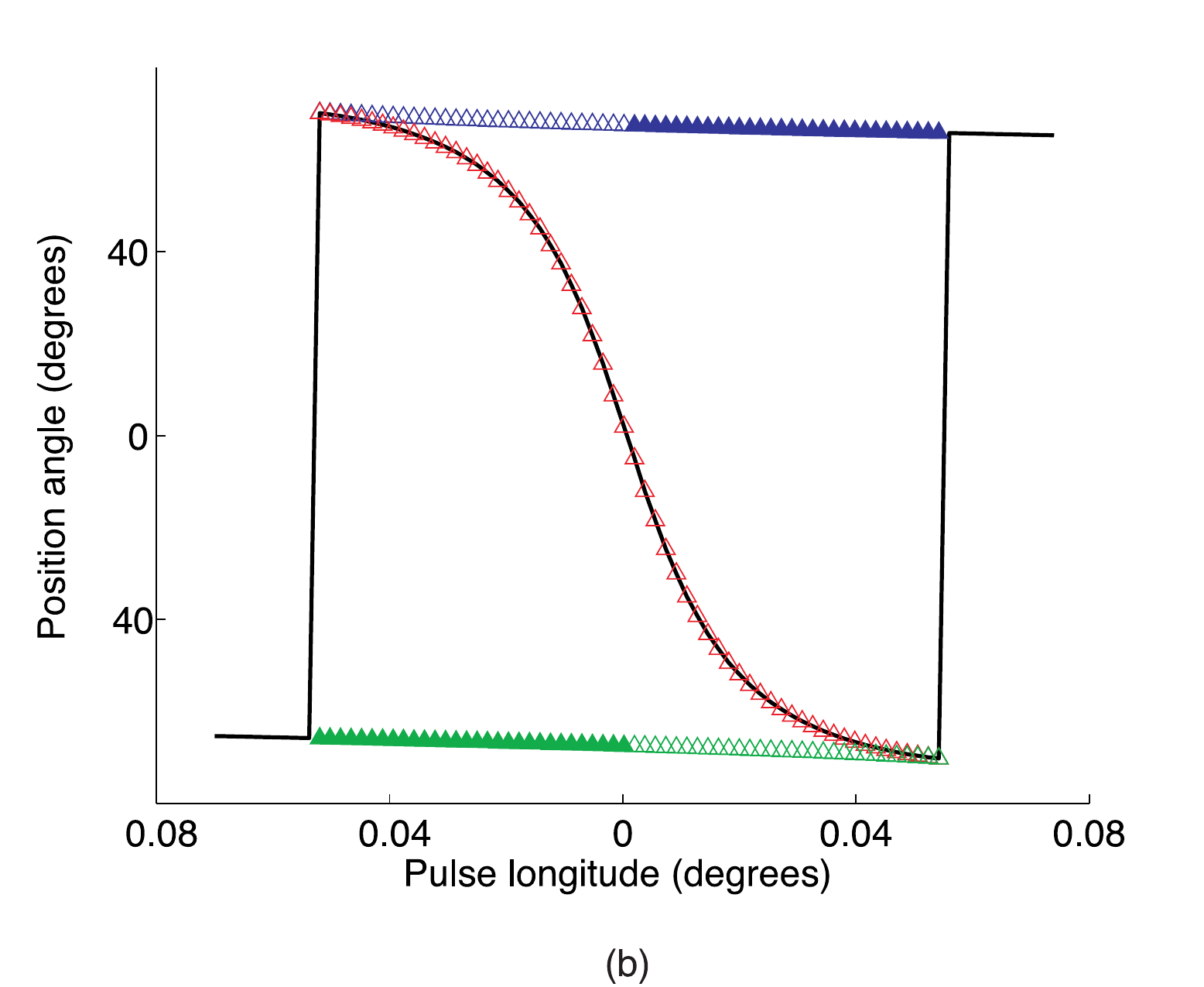}}
\caption{(a) The relative strengths of the three radiation 
modes as observed near the cusp on a sphere of large radius.  
The total field strength (black) and strengths of the 
underlying contributions from the three images of the source 
(green, red, blue) are shown for a source element 
with $r\omega=1.1$ and an observation point that sweeps 
a small arc of the circle $R_P\omega/c=10^{10}$, 
$\theta_P=\pi/2.7$, crossing the envelope near the cusp. 
(b) The corresponding position angles of the contributions 
from the three retarded times (green, red, blue) are shown 
relative to one another and to that of the total field (black); 
the position angles of the dominant contributions are shown 
with open triangles, and those of the weaker contributions 
with filled triangles.}
\label{fig5}
\end{figure}

Field strengths and polarization position angles of the 
three images (radiation modes) close to the cusp 
are shown in Fig.~\ref{fig5}.  Two modes dominate 
everywhere except in the middle of the pulse.  
Moreover, the position angles of two of the modes 
are `orthogonal' and that of the third swings 
across the pulse bridging the other two.  
The constructive interference of the emitted waves 
on the envelope (where two of the contributing 
retarded times coalesce) and on its cusp 
(where all three of the contributing retarded times coalesce) 
gives rise to the divergence of the field of a point-like 
source on these loci~\citep{b9}.  
Here we plot the spatial distribution of 
the field excising the narrow regions in which 
the magnitude of the field exceeds a certain 
threshold~\citep{b13}.

\subsection{The field generated by the entire volume of the source} 
\label{sec:2.3} 

\subsubsection{Nondiffracting subbeams comprising the overall beam and the nonspherical decay of their amplitudes}
\label{sec:2.3.1}

\begin{figure}
\centering
\includegraphics[height=4cm]{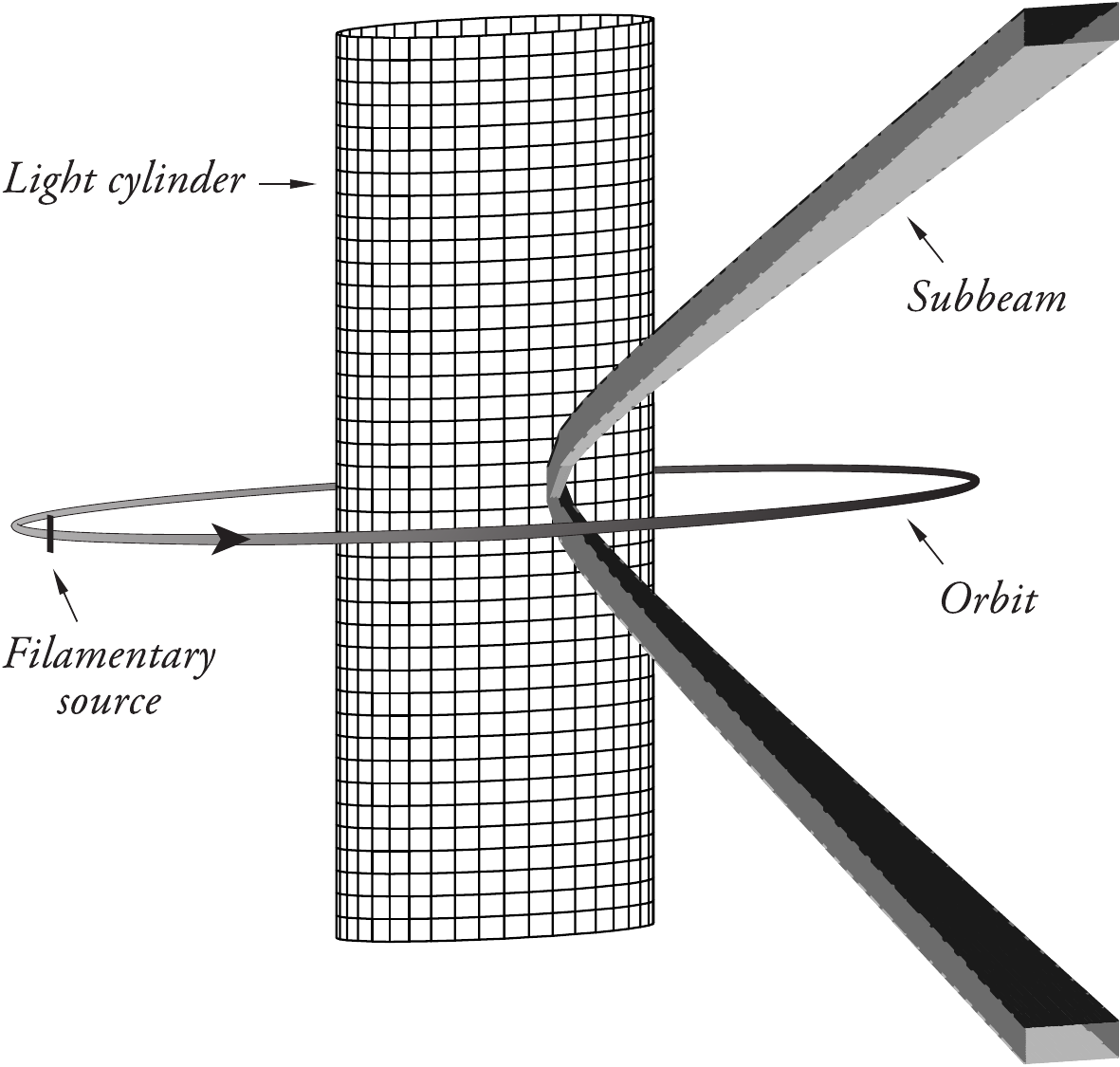}
\caption{Schematic illustration of the light cylinder, 
the filamentary part of the source that approaches 
the observation point with the speed of light and 
zero acceleration at the retarded time, 
the orbit $r=c/(\omega\sin\theta_P)$ of this filamentary source, 
and the subbeam formed by the bundle of cusps that emanate 
from the constituent volume elements of this filament.}
\label{fig6}
\end{figure}

The dominant contribution towards the field of an extended 
source comes from a thin filamentary part of the source that 
approaches the observer, along the radiation direction, 
with the speed of light and zero acceleration at the 
retarded time~\citep{b12}.  For an observation point 
$P$ in the far zone with the coordinates 
$(R_P,\theta_P,\varphi_P)$, this filament is located at 
$r=(c/\omega)\csc\theta_P$, $\varphi=\varphi_P+3\pi/2$ 
and is essentially parallel to the rotation axis (Fig.~\ref{fig6}).  
The collection of cusps of the envelopes of wave fronts that 
emanate from various volume elements of the contributing filament 
form a subbeam whose polar width is nondiffracting: 
the linear dimension of this bundle of cusps in the 
direction parallel to the rotation axis remains the same 
at all distances from the source, so that 
the polar angle $\delta\theta_P$ subtended by the subbeam 
decreases as ${R_P}^{-1}$ with increasing $R_P$ 
[Fig.~\ref{fig6} and~\citet{b12}].  

In that it consists of caustics and so is constantly 
dispersed and reconstructed out of other waves, 
the subbeam in question radically differs from 
a conventional radiation beam [see Appendix D of \citet{b9}].  
The narrowing of its polar width (as ${R_P}^{-1}$) is accompanied 
by a more slowly diminishing intensity (an intensity that 
diminishes as ${R_P}^{-1}$ instead of ${R_P}^{-2}$ with distance), 
so that the flux of energy across its cross sectional area remains 
the same for all $R_P$~\citep{b12}.  This slower rate of decay 
of the emission from a superluminally rotating source has been 
confirmed experimentally~\citep{b6,b7}.

The contributing part of an extended source 
(the filament that approaches the observation 
point with the speed of light and zero acceleration) 
changes as the source rotates (see Fig.~\ref{fig6}).  
In the case of a turbulent plasma with a superluminally 
rotating macroscopic distribution, therefore, 
the overall beam within which the narrow, nonspherically 
decaying radiation is detectable would consist of an 
incoherent superposition of coherent, nondiffracting 
subbeams with widely differing amplitudes and phases 
[similar to the train of giant pulses received from the 
Crab pulsar~\citep{b15}]. 

The overall beam occupies a solid angle whose polar 
and azimuthal extents are independent of the distance $R_P$.  
It is detectable within the polar interval 
$\arccos[(r_<\omega/c)^{-1}]\le\vert\theta_P-\pi/2\vert
\le\arccos[(r_>\omega/c)^{-1}]$, where $[r_<,r_>]$ denotes 
the radial extent of the superluminal part of the source. 
The azimuthal profile of this overall beam reflects the 
distribution of the source density around the cylinder 
$r=c/(\omega\sin\theta_P)$, from which the dominant 
contribution to the radiation arises~\citep{b12}. 

The fact that the observationally-inferred dimensions of 
the plasma structures responsible for the 
emission from pulsars are less than $1$ metre 
in size~\citep{b42} reflects, in the present 
context, the narrowing (as ${R_P}^{-2}$ and ${R_P}^{-3}$, 
respectively) of the radial and azimuthal 
dimensions of the filamentary part of the 
source that approaches the observer with the speed 
of light and zero acceleration at the retarded time~\citep{b12}.
Not only do the nondiffracting subbeams that 
emanate from such filaments account for the nanostructure,
and so the brightness temperature, of the giant 
pulses, but the nonspherical decay of the intensity 
of such subbeams (as ${R_P}^{-1}$ instead 
of ${R_P}^{-2}$) explains why their energy densities 
at their source appear to exceed the energy densities 
of both the plasma and the magnetic field at 
the surface of a neutron star when estimated 
on the basis of the inverse square law \citep{b39}. 

\subsubsection{Frequency spectrum of the radiation by a volume source}
\label{sec:2.3.2}

\begin{figure}
\centering
\subfigure{\includegraphics[width=6cm]{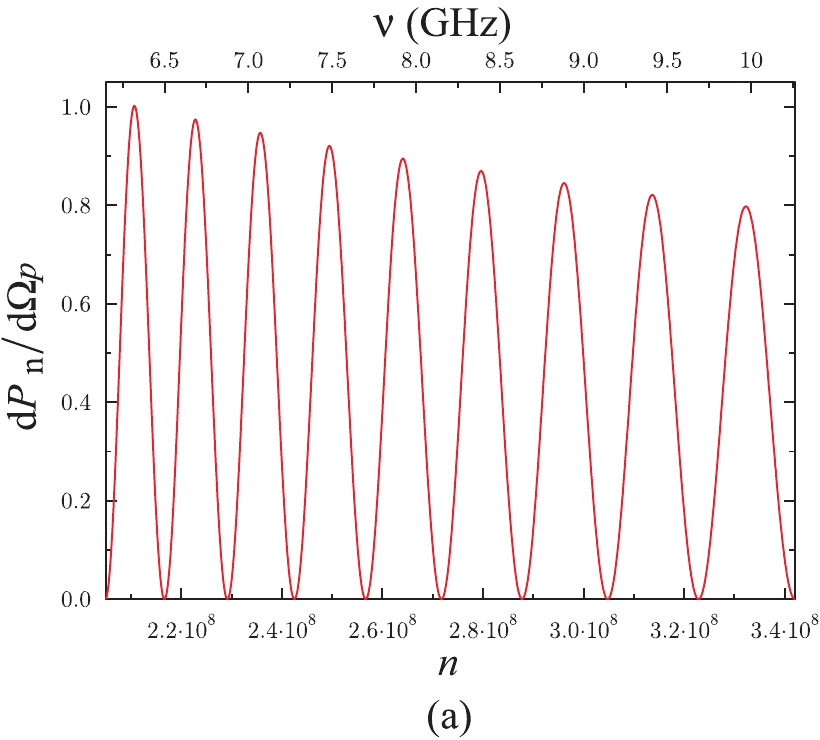}}
\subfigure{\includegraphics[width=6cm]{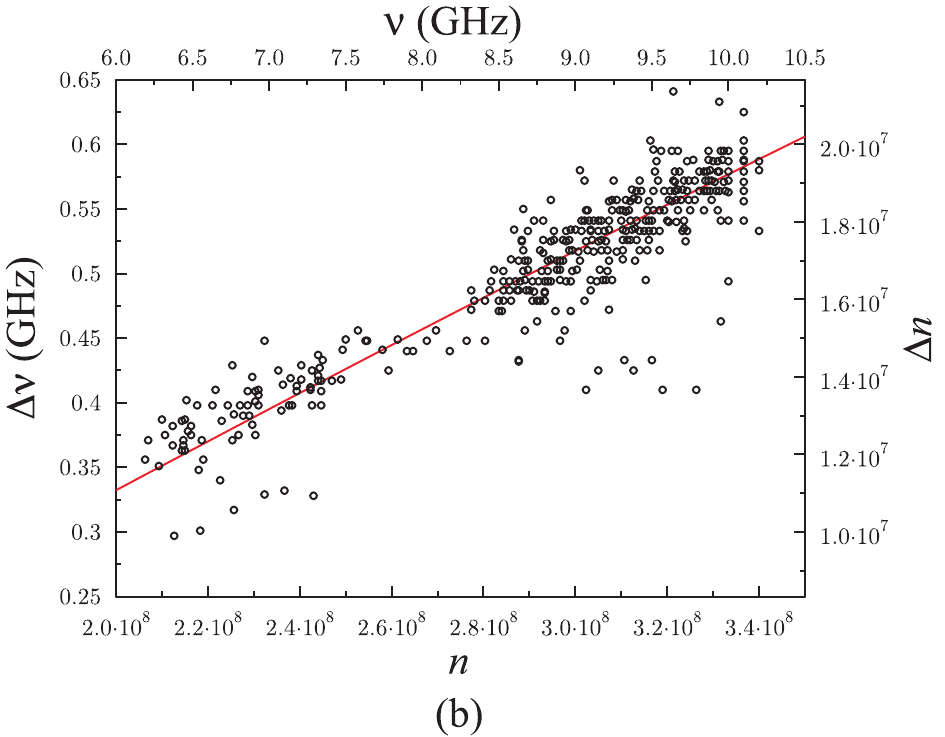}}
\caption{The predicted oscillations of the spectrum of the emission for 
$\omega/(2\pi)=30.3$ Hz and $\Omega/\omega\simeq1.9\times10^4$, shown in (a), have the same spacing as those of the emission bands in the observed spectrum of the Crab pulsar~\citep{b16}, shown in (b) [see~\citet{b1}].}
\label{fig7}
\end{figure}

\begin{figure}
\centering
\includegraphics[height=6cm]{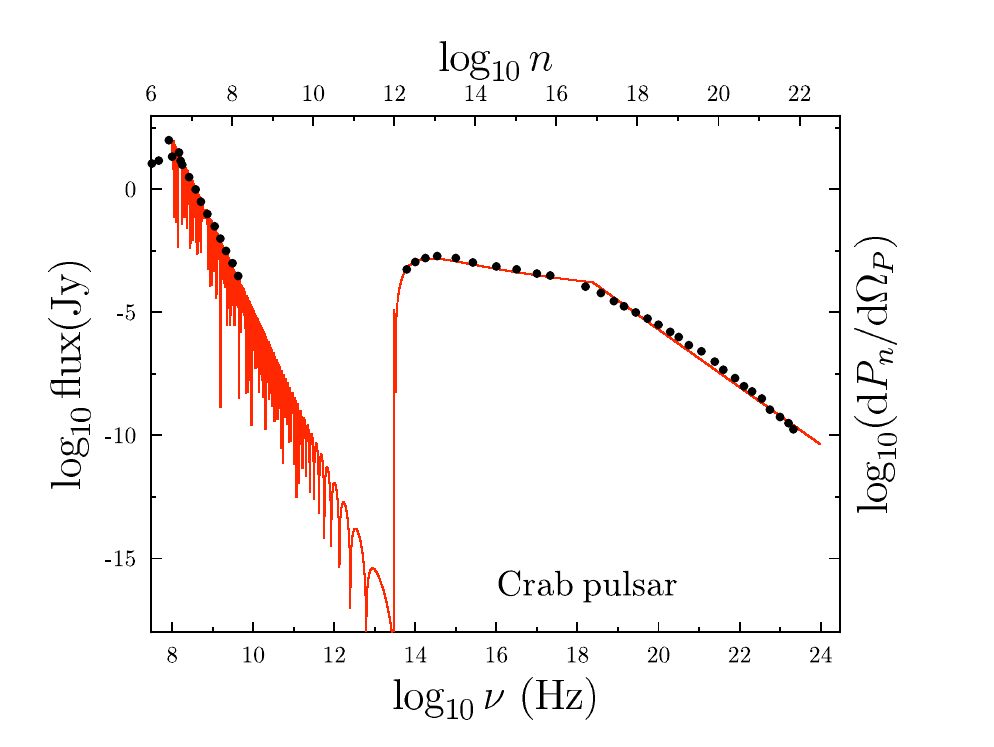}
\caption{The points show observational data (where available) on the spectrum of the Crab pulsar.  The curves show the spectral distribution $\log({\rm d}P_n/{\rm d}\Omega_P)$, predicted by equations (\ref{eq:16here}), (\ref{eq:21}) and (\ref{eq:22}), versus $\log n$ and $\log \nu$ for $\nu=n\omega/(2\pi)\simeq 30.3 n$ Hz and $\Omega/\omega\simeq1.9\times10^4$.  In the model, the recovery of intensity at the ultraviolet peak ($\sim10^{15}$ Hz) is caused by resonant enhancement due to the azimuthal modulation frequency $m\omega/(2\pi)\simeq3\times10^{13}$ Hz.  The steepening of the gradient of the spectrum by $-1$ at $2.4\times10^{18}$ Hz reflects a transition through the Rayleigh distance.}
\label{fig8}
\end{figure}

The spectrum of the radiation emitted by an extended 
source with a superluminally rotating distribution pattern
is oscillatory with oscillations whose spacing increase 
with frequency [see~\citet{b1,b10} 
and equation~(\ref{eq:16here}) below].  While the Bessel 
function describing synchrotron radiation has an 
argument smaller than its order and so decays 
exponentially with increasing frequency, the 
Bessel function appearing in the present analysis~\citep{b10}, 
whose argument exceeds its order, is an 
oscillatory function of frequency with an 
amplitude that decays only algebraically~\citep{b19}. 
Figure~\ref{fig7} shows that the spacing 
of the emission bands in the spectrum of 
the Crab pulsar fit the predicted oscillations 
for an appropriate choice of the single 
parameter $\Omega/\omega$.  The value of 
this parameter, thus implied by the data of~\citet{b16}, 
places the last peak of the oscillating spectrum at a 
frequency $(\sim\Omega^3/\omega^2)$ that agrees with the 
position of the ultraviolet peak in the spectrum of the 
Crab pulsar.  By inferring the remaining free parameter $m$ in 
equation (\ref{eq:3}) from the observational data and by adjusting the 
spectral indices of the power laws describing the spectral 
decomposition of the source density ${\bf s}$ in various
frequency bands [see Section~(\ref{sec:4})], 
we are thus able to account for the continuum spectrum of the Crab 
pulsar over 16 orders of magnitude of frequency (Fig.~\ref{fig8}).

At radiation frequencies higher than $\sim10^{18}$ Hz,
the Rayleigh distance [see Section~\ref{raysec} and~\citet{b919}] 
becomes larger than
the distance of the Crab pulsar from the Earth.  
For an observer within the Rayleigh distance, the emission arises from a 
narrower radial extent of the source, and so has a higher 
degree of mean polarization, the higher the frequency.  
The degree of circular polarization of such a high-frequency 
emission decreases with increasing frequency $\nu$ as $\nu^{-1/3}$, 
so that this emission is essentially 100 per cent linearly 
polarized at all phases, including off-pulse phases 
[see equation~(\ref{eq:a24}) below]. Note that, in this model, 
the intensity and polarization of the off-pulse emission, too, 
reflect the distributions of density and orientation of the 
emitting current around the cylinder $r=c/(\omega\sin\theta_P)$ 
within the pulsar magnetosphere~\citep{b12}.

We will see in Appendix A that the reason the spectrum of the 
emission from the source distribution 
(\ref{eq:3}) is not limited, in the superluminal case, 
to just the frequencies $\Omega\pm m\omega$ is the following.
On the one hand, each element ${\hat\varphi}$ of a constituent 
ring of the source makes its contribution towards the observed 
field at the discrete set of retarded times 
$\omega t=\varphi_P-{\hat\varphi}+3\pi/2+2k\pi$ 
at which it approaches the observer with the speed of light 
and zero acceleration, i.e.\ {\em periodically} with the 
period $2\pi/\omega$ ($k$ is an integer).  On the other hand, 
temporal modulations of 
the density of this source occur on a time scale $2\pi/\Omega$ 
that is incommensurable with the period $2\pi/\omega$.  
Because the emission time of the element labelled by ${\hat\varphi}$ 
is fixed (via $\omega t=\varphi_P-{\hat\varphi}+3\pi/2+2k\pi$) 
by its initial azimuthal position (${\hat\varphi}$), the temporal 
[$\cos(\Omega t)$] and spatial [$\cos(m{\hat\varphi})$] 
modulations of source (\ref{eq:3}) effectively combine 
into a single variation [$\propto\exp({\rm i}\Omega{\hat\varphi}/\omega)
\cos(m{\hat\varphi})$].  It is the Fourier decomposition of 
this variation with ${\hat\varphi}$, or equivalently with
the retarded time $t$, that together with the incommensurablity 
of the two periods ($2\pi/\omega$ and $2\pi/\Omega$), and 
the finiteness of the domain of definition of ${\hat\varphi}$, 
results in a spectrum containing all frequencies 
[see Fig.\ 5 of~\citet{b10}].

\section{Theoretical fits to the multiwavelength data on pulsar spectra} 
\label{sec:4}
\subsection{The equation used to create the fits}

Using the result derived in Appendix A [equation~\ref{eq:a28})],
we can write the frequency dependence of the power that 
is radiated by the source distribution 
(\ref{eq:3}) per harmonic per unit solid angle as
\begin{eqnarray}
\frac{{\rm d}P_n}{{\rm d}\Omega_P}& \propto &{S_1(n)}^2{\rm Ai}^2\left[-\left(\frac{2}{n}\right)^{1/3}\frac{\Omega}{\omega}\right]\nonumber\\
&&\mbox{}+{S_2(n)}^2\left(\frac{2}{n}\right)^{2/3}{\rm Ai^\prime}^2\left[-\left(\frac{2}{n}\right)^{1/3}\frac{\Omega}{\omega}\right]\nonumber\\
&&\mbox{}+2S_3(n)^2\left(\frac{2}{n}\right)^{1/3}{\rm Ai}\left[-\left(\frac{2}{n}\right)^{1/3}\frac{\Omega}{\omega}\right]\nonumber\\
&&\mbox{}\times{\rm Ai^\prime}\left[-\left(\frac{2}{n}\right)^{1/3}\frac{\Omega}{\omega}\right],
\label{eq:16here}
\end{eqnarray}
in which
\[
S_1(n)=n^{2/3}\vert K_r\vert K_{\varphi_0}\left(\vert{\bar s}_r\vert^2+\vert{\bar s}_\varphi\cos\theta_P-{\bar s}_z\sin\theta_P\vert^2\right)^{1/2},
\]
\[
S_2(n)=n^{2/3}\vert K_r\vert K_{\varphi_0}\left(\vert{\bar s}_\varphi\vert^2+\vert{\bar s}_r\vert^2\cos^2\theta_P\right)^{1/2},
\]
and
\begin{eqnarray*}
S_3(n)& = &n^{2/3}\vert K_r\vert K_{\varphi_0}\nonumber\\
&&\times \left\{\Im\left[{{\bar s}_r}^*\cos\theta_P\left({\bar s}_\varphi\cos\theta_P-{\bar s}_z\sin\theta_P\right)-{\bar s}_r{{\bar s}_\varphi}^*\right] \right\}^{1/2}.
\end{eqnarray*}
Here, ${\bar s}_{r,\varphi,z}$ are the Fourier components of 
the source densities $s_{r,\varphi,z}\vert_{{\hat r}=\csc\theta_P}$
with respect to $z$ [see equation~(\ref{eq:a25})],
$\Im$ and the superscript star denote the imaginary 
part and the conjugate of a complex variable, respectively, and
$(R_P,\theta_P,\varphi_P)$ are the spherical 
polar coordinates of the observation point 
$P$.
The function $K_{\varphi_0}$ is defined by
\begin{equation}
K_{\varphi_0}=(-1)^{n+m}\sin\left(\frac{\pi\Omega}{\omega}\right)\left(\frac{{\mu_+}}{n-\mu_+}+\frac{{\mu_-}}{n-\mu_-}\right),
\label{qdef}
\end{equation}
and the various frequencies involved ($n$, $\mu_+$, etc.)
are tabulated for convenience in
Table~\ref{table1}. 
Finally, the function $K_r$ is defined for various frequency ranges
in Table~\ref{table2};
${\hat r}_<$ and ${\hat r}_>$ denote the radial boundaries of
the part of the source that contributes towards 
the radiation at $P$ in units of the light-cylinder radius
$c/\omega$.
\begin{table}
\caption{Frequencies involved in the fitting of the pulsar spectra.
For convenience in comparisons with observations,
most parameters are presented in frequency units (i.e.\ in Hz)
rather than as angular frequencies (i.e.\ in radians per second).}
\centering
\begin{tabular}{|c|c|}
\hline
Frequency & Definition \\
\hline
$\frac{\omega}{2\pi}$ & pulsar rotation frequency \\
\hline
$\nu=\frac{n\omega}{2\pi}$ & observation frequency\\
\hline
$\frac{m\omega}{2\pi}$ & azimuthal modulation frequency\\
\hline
$\frac{\Omega}{2 \pi}$ & temporal modulation frequency\\
\hline
$\mu_+$ & $\frac{\Omega}{\omega}+m$ \\
\hline
$\mu_-$ & $\frac{\Omega}{\omega}-m$ \\
\hline
\label{table1} 
\end{tabular} 
\end{table}

\begin{table}
\caption{Limiting values of parameters and functions in 
various frequency ranges; 
the harmonic number $n$ yields the observation frequency 
$\nu=n\omega/2\pi$ in units of the rotation frequency $\omega/2\pi$.} 
\centering 
\begin{tabular}{|c|c|c|c|} 
\hline 
Function & Frequency range & Limiting value \\ 
\hline 
${\rm Ai}\left[-\left(\frac{2}{n}\right)^{1/3}\frac{\Omega}{\omega}\right]$ & $n\ll \left(\frac{\Omega}{\omega}\right)^3$ & 
$\left(\frac{2}{n}\right)^{-\frac{1}{12}}\left(\frac{\Omega}{\omega}\right)^{-\frac{1}{4}}$\\
\hline
${\rm Ai}\left[-\left(\frac{2}{n}\right)^{1/3}\frac{\Omega}{\omega}\right]$ & $n\gg \left(\frac{\Omega}{\omega}\right)^3$ & 
${\rm Ai}(0)$\\
\hline
${\rm Ai}^{\prime}\left[-\left(\frac{2}{n}\right)^{1/3}\frac{\Omega}{\omega}\right]$ & $n\ll \left(\frac{\Omega}{\omega}\right)^3$ & 
$\left(\frac{2}{n}\right)^{\frac{1}{12}}\left(\frac{\Omega}{\omega}\right)^{\frac{1}{4}}$\\
\hline
${\rm Ai}^{\prime} \left[-\left(\frac{2}{n}\right)^{1/3}\frac{\Omega}{\omega}\right]$ &  $n\gg \left(\frac{\Omega}{\omega}\right)^3$ & 
${\rm Ai}^{\prime}(0)$\\
\hline
$K_{\varphi_0}$ & $n\ll \mu_+$ & $\sim 1$ \\
\hline
$K_{\varphi_0}$ & $n\sim m \gg \frac{\Omega}{\omega}$ & $\sim n$ \\
\hline
$K_{\varphi_0}$ & $n \gg \mu_+$ & $\sim n^{-1}$ \\
\hline 
$K_r$ & $n\ll \frac{\pi{\hat R}_P}{({\hat r}_>-{\hat r}_<)^2}$ & $\simeq{\hat r}_>-{\hat r}_<$ \\
\hline
$K_r$ & $n\gg \frac{\pi{\hat R}_P}{({\hat r}_>-{\hat r}_<)^2}$ & 
$\simeq\left(\frac{2\pi}{n}{\hat R}_P\right)^{\frac{1}{2}}\exp\left(-\frac{{\rm i}\pi}{4}\right)$\\
\hline
\label{table2} 
\end{tabular} 
\end{table}

\subsection{Fit procedure} 
The pulsar spectra are fitted to the predicted spectrum
(\ref{eq:16here}) by adjusting the modulation frequencies 
$\Omega/2\pi$ and $m\omega/2\pi$ of the source (see Table~\ref{table1}), 
and by specifying the required 
rates of change of the Fourier components ${\bar s}_{r,\varphi,z}$ 
of the source density (or equivalently,
and more practically, $S_1$, $S_2$ and $S_3$) 
with frequency $\nu=n\omega/2\pi$.  
Our objective in each case is to illustrate
that the observed spectrum can be faithfully represented by 
that in equation~(\ref{eq:16here}) within a wide margin of accuracy.
We will model the frequency dependence of 
the factors that multiply the Airy functions in 
this equation by as simple a set of power laws 
as allowed by the available data. We will also
extrapolate the spectra thus derived into the 
frequency intervals for which no data are available.

The position of the peak of the last oscillation in the spectrum of 
the radiation described by equation (\ref{eq:16here}) is determined by
the ratio $\Omega/\omega$. 
Observational data  
specify the value of this ratio for each of the 9 pulsars we 
consider quite accurately [see also~\citet{b1}]. 
However, the existing data 
are insufficient to determine the remaining parameters of the 
source [frequency $m\omega$ of the azimuthal 
modulations of the 
current distribution (\ref{eq:3}) and the frequency dependence of the 
Fourier components 
${\bar s}_{r,\varphi,z}$ of the amplitude of this current] uniquely.  
We will assume that the amplitudes 
${\bar s}_{r,\varphi,z}$ [or equivalently the functions $S_i(n)$, $i=1,2,3$,
that appear in equation (\ref{eq:16here})] have power-law dependences of the
form $n^{\alpha_i}$ on the harmonic number $n$ with differing exponents 
$\alpha_i$ in different
frequency bands.  The relative amplitudes, $S_2/S_1$ and $S_3/S_2$, and 
the values of $\alpha_i$ are then inferred from the data. 

We start plotting each figure 
with the inferred values of 
the above adjustable parameters in the radio band and continue 
the resulting theoretical curves as far as the next
patch of available data for which one or more of these parameters have 
different values. In a few cases (Vela, Geminga and B0656+14), we
will cover an interval for which no data are available by 
extrapolating the theoretical curves implied by the data on opposite
sides of the empty interval until they meet. 
We are not predicting that the spectrum would necessarily follow the
plotted curves in frequency intervals where no data are available.  
Not only is the polarization current distribution in the magnetosphere
of a pulsar not expected to have such a simple space-time 
structure as that described by 
equation (\ref{eq:3}), but even in such a simplified description, 
the amplitudes ${\bar s}_{r,\varphi,z}$ need not have the power-law 
forms we have approximated them by.  

Before embarking on a detailed description of the fits,
we show, in the following three Subsections,
how the behaviour of the various components
of equation~(\ref{eq:16here}) (see Table~\ref{table2})
leads to some ``universal''
features of the emission spectra of pulsars.
Evidently, many of the gross phenomena seen in
pulsar observations are {\it quite general consequences}
of the fact that they arise from a rotating, superluminal source.
It is only in modelling the {\it detailed behaviour} of the spectra
that it is necessary to
specify parameters $S_1$, $S_2$ and $S_3$
(which represent intricacies of the
`weather' around the emitting region as opposed to the 
`climate' in the pulsar magnetosphere) with any accuracy.
\subsection{Resonant enhancement}
According to equation~(\ref{qdef}), the value of 
$K_{\varphi_0}$ (and, hence, those of $S_1$ , $S_2$ and $S_3$)
can undergo a sudden change as a result of the resonance
between the radiation frequency $\omega n/2\pi$ and one of the source 
frequencies $\omega \mu_\pm/2\pi$.  
As summarized in Table~\ref{table2},
$K_{\varphi_0}$  is independent of $n$ ($\sim1$) when $n\ll\mu_+$, 
decays as $n^{-1}$ when $n\gg\mu_+$, and is of the order 
of $n$ when $n\sim m\gg\Omega/\omega$ 
[see equation(\ref{qdef})]; hence, its value sharply increases by 
several orders of magnitude for $n\sim m\gg\Omega/\omega$. 
This is the primary cause of the
recovery of the intensity  
at higher frequencies that will be seen
in the fitted spectra below.
\subsection{The Airy function and its derivative}

The Anger functions appearing in the general expression 
for the radiated power [equation (66) of~\citet{b10}] reduce 
to Bessel functions for $n\gg1$ and to Airy functions for 
$n\gg\Omega/\omega$ [see Appendix C of \citet{b1}].  
In most parts of the spectra modelled 
here, $\Omega/\omega\ll n \ll (\Omega/\omega)^3$ 
and so ${\rm Ai'/Ai}$ is greater than unity
(Table~\ref{table2}).
Consequently, when the second term of
equation~(\ref{eq:16here}) is employed in the fits below, 
$S_2/S_1 $ is given values of the order of $10^6$.  
In other words, where invoked, the second term is actually
comparable to the first term everywhere apart from the region past the
last maximum.  The dependence on $n$ of band spacings 
and amplitudes of the two functions 
${\rm Ai}$ and
$\rm Ai^{\prime}$ are sufficiently different for 
the inclusion of the second term to
make a difference to the shape of the spectrum.

The third term of equation~(\ref{eq:16here}) declines with $n$
at a similar rate to the second term in regions
of the spectrum where $n << (\Omega/\omega)^3$.
However, it was not found necessary to invoke this
term in the fits below, probably because
$S_3$ is much smaller
than $S_1$ and $S_2$ when the phases of the Fourier components 
$({\bar s}_r,{\bar s}_\varphi,{\bar s}_z)$ of the source 
density ${\bf s}$ do not bear any relationship with one another.
Unlike $S_1$ and $S_2$
that depend on the absolute values of these (generally complex)
quantities, $S_3$ sensitively depends on their relative phases.  In
addition, the first two terms are positive for all $n$, while the third term
oscillates between positive and negative values.  Since the measurement of
this term by any instrument would entail an integration with respect to
frequency over a non-zero bandwidth, these oscillations 
will likely make the detected
value of the third term negligibly smaller than those of $S_1$ and $S_2$.
\subsection{The Rayleigh distance}
\label{raysec}
The coefficient 
$K_r$ changes from being independent of 
$n$ to decaying as $n^{-1/2}$ when 
$n$ increases past $\pi{\hat R}_P/({\hat r}_>-{\hat r}_<)^2$ (Table~\ref{table2}).
On rearranging this condition for $n$ in terms of $R_P$, we obtain
\begin{equation}
R_P \sim \frac{a^2}{\lambda} \equiv d_{\rm R},
\end{equation}
where $\lambda=c/\nu$ is the wavelength of the electromagnetic radiation  
and $a=r_>-r_<$ is the radial extent of the emitting current.
In geometrical optics,
the parameter $d_{\rm R}$ is known as the {\it Rayleigh 
distance} for an aperture or source
of linear dimension $a$; it corresponds roughly
to the transition between the conditions for 
Fresnel and Fraunhofer diffraction~\citep{b919}.

Whilst the Rayleigh distance
is most frequently encountered in optical or radio
experiments when using monochromatic radiation
and varying the distance between source and observer~\citep{b919},
here we cross it by keeping $R_P$ fixed and varying the
frequency $\nu=n\omega/2\pi$.
In the following fits to data,
the transition across 
the Rayleigh distance accounts for the steepening of the 
observed spectra
by the factor $n^{-1}$ in the X-ray band
and enables us to derive the 
radial extent ${\hat r}_>-{\hat r}_<$ 
of the emitting plasma that contributes towards the 
subpulse detected at the observation point.

Having described the general behaviour of equation~(\ref{eq:16here}),
we now embark on the fits of this equation to observational data.

\subsection{The Crab pulsar, PSR B0531+21}
\label{sec:4.1}

We have taken the phase-averaged spectra of the pulsed 
emission that is received from the Crab pulsar in various frequency bands 
(the data points shown in Fig.~\ref{fig8}) 
from~\citet{b17}.  The curve in Fig.~\ref{fig8}
represents the radiation flux given by equation (\ref{eq:16here})
for the following values of the parameters: 

\begin{equation}
\frac{\Omega}{\omega}\simeq1.9\times10^4\quad {\rm with}\quad \frac{\omega}{2\pi}=30.3\, {\rm Hz},
\label{eq:20}
\end{equation}

\begin{eqnarray}
S_1\propto\left\{ \begin{array}{lrll}
n^{-11/6}     & 3.2\times 10^6 &< n <&  10^{12}\\
n^{-1/12}      & 10^{12} &< n <&  7.9\times 10^{16}\\
n^{-7/12}     & 7.9\times 10^{16} &< n <& 3.2\times 10^{22},
\end{array}\right.
\label{eq:21}
\end{eqnarray}

\begin{equation}
\frac{S_2}{S_1}\ll1,\quad\frac{S_3}{S_1}\ll1
\label{eq:22}
\end{equation}
for all $n$, and $m=10^{12}$.

Given that the Crab pulsar is at the distance 
$R_P\simeq 6.2\times10^{21}$~cm~\citep{b22} and has a 
light cylinder with the radius 
$c/\omega= 1.6\times10^8$~cm, 
the transition through the Rayleigh distance
would account for the observed 
steepening of its spectrum at $n=7.9\times10^{16}$ 
(Fig.~\ref{fig8}) if the radial extent ${\hat r}_>-{\hat r}_<$ 
of the emitting plasma is a fraction $3.9\times10^{-2}$ of the 
light-cylinder radius.

\subsection{The Vela pulsar, PSR B0833-45}
\label{sec:4.2}

\begin{figure}
\centering
\includegraphics[height=6cm]{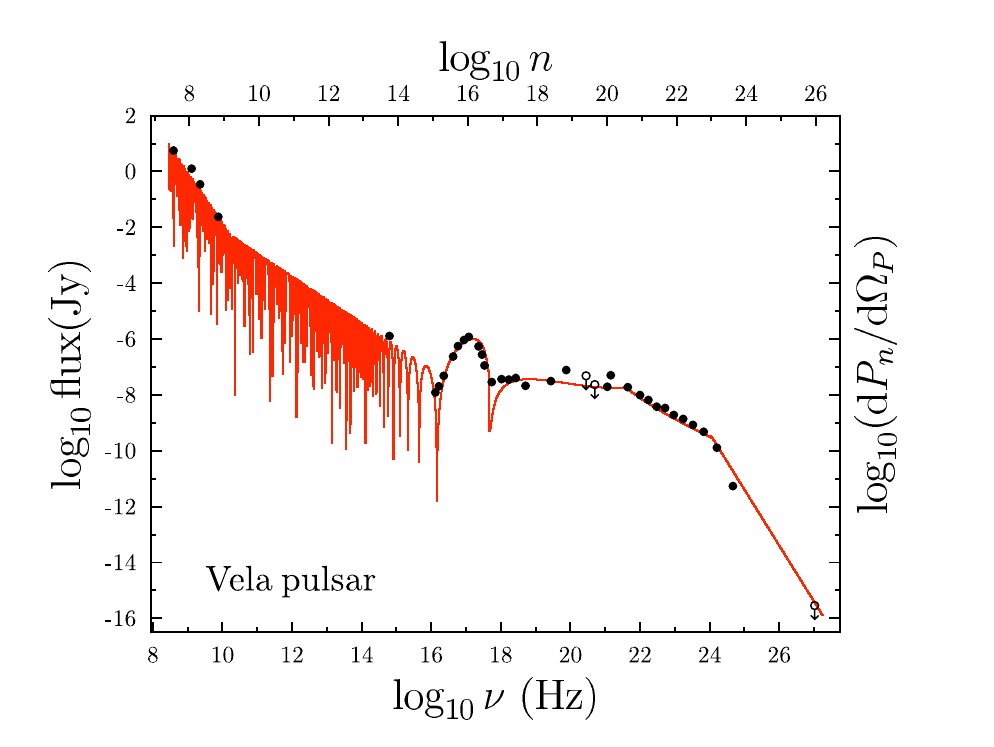}
\caption{The points show observational data (where available) from the spectrum of the Vela pulsar.  The curves show the spectral distribution $\log({\rm d}P_n/{\rm d}\Omega_P)$, predicted by equations (\ref{eq:16here}), (\ref{eq:24}) and (\ref{eq:25}), versus $\log n$ and $\log \nu$ for $\nu=n\omega/(2\pi)\simeq 11.2 n$ Hz and $\Omega/\omega\simeq2.8\times10^5$.  In the model, the recovery of intensity in the $X$-ray band is caused by resonant enhancement due to the azimuthal modulation frequency $m\omega/(2\pi)\simeq1.5\times10^{16}$ Hz.  The steepening of the gradient of the spectrum by $-1$ at $3.6\times10^{21}$ Hz reflects the transition through the Rayleigh distance.}
\label{fig9}
\end{figure}

We have taken the data points shown in Fig.~\ref{fig9}
from~\citet{b21} [see also~\citet{b17}].  The curve in Fig.~\ref{fig9}
represents the radiation flux given by equation (\ref{eq:16here})
for the following values of the parameters: 

\begin{equation}
\frac{\Omega}{\omega}\simeq2.82\times10^5\quad {\rm with}\quad \frac{\omega}{2\pi}=11.2\, {\rm Hz},
\label{eq:23}
\end{equation}

\begin{eqnarray}
S_1\propto\left\{ \begin{array}{lrll}
n^{-1/2}     & 2.5\times 10^7 &< n <&  1.3\times 10^9\\
n^0      & 1.3\times 10^9 &< n <&  1.3\times 10^{15}\\
n^{3/2}     & 1.3\times 10^{15} &< n <& 4\times 10^{16}\\
n^{1/5}     & 4\times 10^{16} &< n <& 3.2\times 10^{20}\\
n^{-3/10}     & 3.2\times 10^{20} &< n <& 10^{23}\\
n^{-1}     & 10^{23} &< n <& 1.6\times 10^{26},
\end{array}\right.
\label{eq:24}
\end{eqnarray}

\begin{equation}
\frac{S_2}{S_1}=10^7,\quad\frac{S_3}{S_1}\ll1
\label{eq:25}
\end{equation}
for all $n$, and $m=1.3\times10^{15}$.

Given that the Vela pulsar is at the distance 
$R_P\simeq 1.5\times10^{21}$~cm~\citep{b22} and has a 
light cylinder with the radius 
$c/\omega= 4.3\times10^8$~cm, 
the transition through the Rayleigh distance 
would account for the observed 
steepening of its spectrum at $n=3.2\times10^{20}$ 
(Fig.~\ref{fig9}) if the radial extent ${\hat r}_>-{\hat r}_<$ 
of the emitting plasma is a fraction $1.9\times10^{-4}$ of the 
light-cylinder radius.    

\subsection{The Geminga pulsar, PSR J0633+1746}
\label{sec:4.3}

\begin{figure}
\centering
\includegraphics[height=6cm]{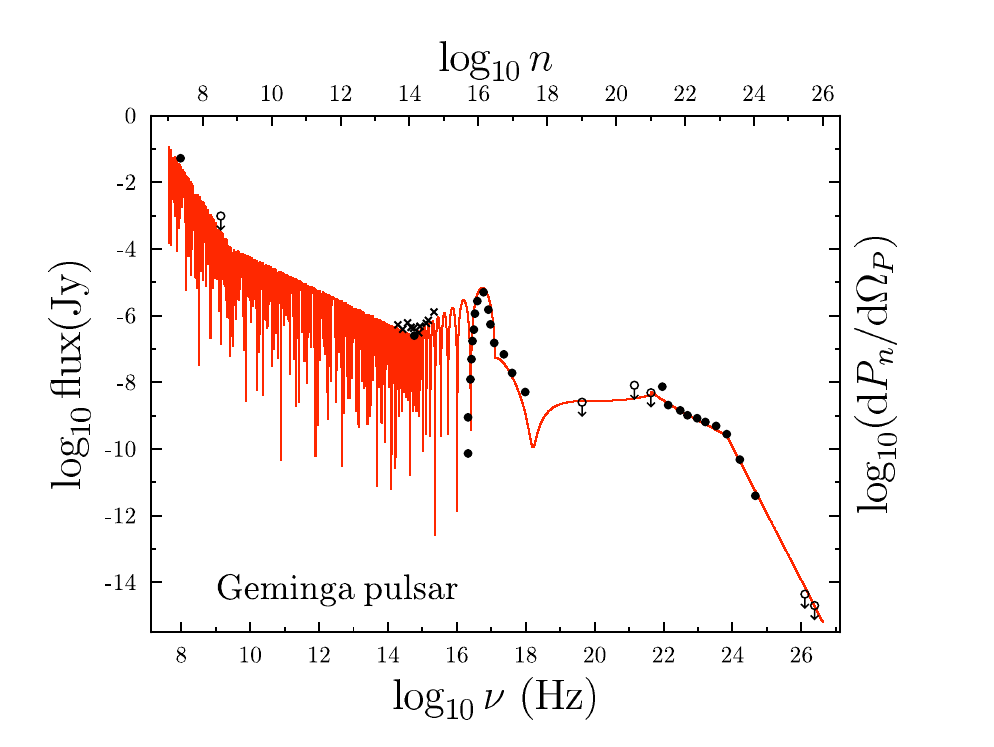}
\caption{The points show observational data (where available) from the spectrum of the Geminga pulsar.  The curves show the spectral distribution $\log({\rm d}P_n/{\rm d}\Omega_P)$, predicted by equations (\ref{eq:16here}), (\ref{eq:27}) and (\ref{eq:28}), versus $\log n$ and $\log \nu$ for $\nu=n\omega/(2\pi)\simeq 4.22 n$ Hz and $\Omega/\omega\simeq5.89\times10^5$.  In the model, the recovery of intensity over the optical and $X$-ray bands is caused by resonant enhancement due to the azimuthal modulation frequency $m\omega/(2\pi)\simeq6.7\times10^{14}$ Hz.  The steepening of the gradient of the spectrum by $-1$ at $4.2\times10^{21}$ Hz reflects the crossing of the Rayleigh distance.}
\label{fig10}
\end{figure}

We have taken the data points shown in Fig.~\ref{fig10} from
\citet{b21} and~\citet{b23} [see also~\citet{b17}]. 
 The curve in Fig.~\ref{fig10}
represents the radiation flux given by equation (\ref{eq:16here})
for the following values of the parameters: 

\begin{equation}
\frac{\Omega}{\omega}\simeq5.89\times10^5\quad {\rm with}\quad \frac{\omega}{2\pi}=4.22\, {\rm Hz},
\label{eq:26}
\end{equation}

\begin{eqnarray}
S_1\propto\left\{ \begin{array}{lrll}
n^{-5/12}     & 10^7 &< n <& 1.58\times 10^9\\
n^{1/6}       & 1.6\times10^9 &< n <& 1.6\times 10^{14}\\
n^{4/15}       & 1.6\times10^{14} &< n <& 3\times 10^{16}\\
n^{1/4}      & 3\times 10^{16} &< n <& 10^{21}\\
n^{-1/4}      & 10^{21} &< n <& 1.8\times10^{23}\\
n^{-1}      & 1.8\times10^{23} &< n <& 10^{26},
\end{array}\right.
\label{eq:27}
\end{eqnarray}

\begin{eqnarray}
\frac{S_2}{S_1}=\left\{ \begin{array}{lc}
0            & 1.6\times10^{14} < n < 3\times 10^{16}\\
10^7              & \mbox{otherwise,}
\end{array}\right.
\label{eq:28}
\end{eqnarray}
$S_3\ll S_1$ for all $n$, and $m=1.6\times10^{14}$.

Given that the Geminga pulsar is at the distance 
$R_P\simeq 4.6\times10^{20}$~cm~\citep{b22} and has a 
light cylinder with the radius 
$c/\omega= 1.1\times10^9$~cm, 
crossing the Rayleigh distance
would account for the observed 
steepening of its spectrum at $n=10^{21}$ 
(Fig.~\ref{fig10}) if the radial extent ${\hat r}_>-{\hat r}_<$ 
of the emitting plasma is a fraction $3.6\times10^{-5}$ 
of the light-cylinder radius.  

\subsection{PSR B0656+14}
\label{sec:4.8}

\begin{figure}
\centering
\includegraphics[height=6cm]{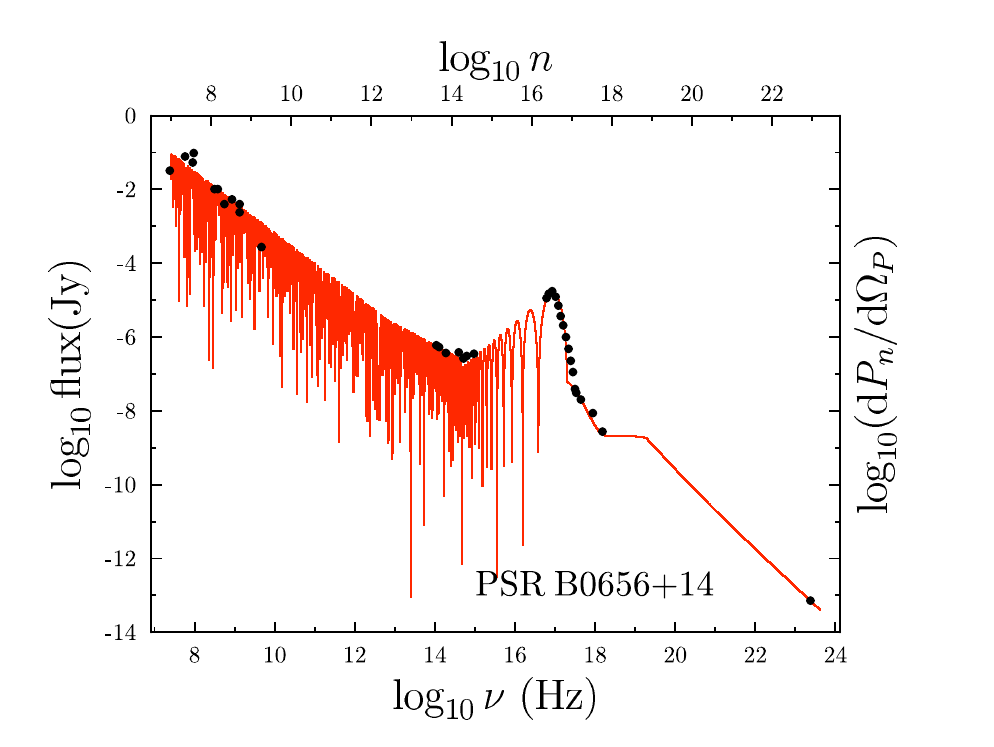}
\caption{The points show observational data (where available) from the spectrum of B0656+14.  The curves show the spectral distribution $\log({\rm d}P_n/{\rm d}\Omega_P)$, predicted by equations (\ref{eq:16here}), (\ref{eq:43}) and (\ref{eq:44}), versus $\log n$ and $\log \nu$ for $\nu=n\omega/(2\pi)\simeq 2.6 n$ Hz and $\Omega/\omega\simeq5.89\times10^5$.  In the model, the recovery of intensity over the optical and $X$-ray bands is caused by resonant enhancement due to the azimuthal modulation frequency $m\omega/(2\pi)\simeq5.2\times10^{14}$ Hz.  The steepening of the gradient of the spectrum by $-1$ at $1.8\times10^{19}$~Hz corresponds to crossing the Rayleigh distance.}
\label{fig11}
\end{figure}

We have taken the data points shown in Fig.~\ref{fig11} from
\citet{b25},~\citet{b26}, and~\citet{b27}. 
The curve in Fig.~\ref{fig11}
represents the radiation flux given by equation (\ref{eq:16here})
for the following values of the parameters: 

\begin{equation}
\frac{\Omega}{\omega}\simeq7.94\times10^5\quad {\rm with}\quad \frac{\omega}{2\pi}=2.6\, {\rm Hz},
\label{eq:42}
\end{equation}

\begin{eqnarray}
S_1\propto\left\{ \begin{array}{lrll}
n^{-1/2}     & 10^7& < n <& 3.2\times10^{12}\\
n^{-3/8}      & 3.2\times10^{12}&< n <& 2\times 10^{14}\\
n^{3/8}     & 2\times10^{14}&< n <& 7.6\times10^{16}\\
n^0      & 7.6\times10^{16}&< n <& 7.1\times 10^{18}\\
n^{-1/2}     & 7.1\times10^{18}&< n <& 1.58\times 10^{23},
\end{array}\right.
\label{eq:43}
\end{eqnarray}
\begin{eqnarray}
\frac{S_2}{S_1}=\left\{ \begin{array}{lc}
0            & 10^7 < n < 7.6\times 10^{16}\\
1.78\times10^6              & \mbox{otherwise,}
\end{array}\right.
\label{eq:44}
\end{eqnarray}
$S_3\ll S_1$ for all $n$, and $m=2\times10^{14}$.

Given that PSR B0656+14 is at the distance 
$R_P\simeq 2.3\times10^{21}$~cm~\citep{b22} and has a 
light cylinder with the radius 
$c/\omega= 1.8\times10^9$~cm, 
the transition through the Rayleigh distance
would account for the observed 
steepening of its spectrum at $n=7.1\times10^{18}$ 
(Fig.~\ref{fig11}) if the radial extent ${\hat r}_>-{\hat r}_<$ 
of the emitting plasma is a fraction $7.5\times10^{-4}$ 
of the light-cylinder radius.   
   
\subsection{PSR B1055-52}
\label{sec:4.5}

\begin{figure}
\centering
\includegraphics[height=6cm]{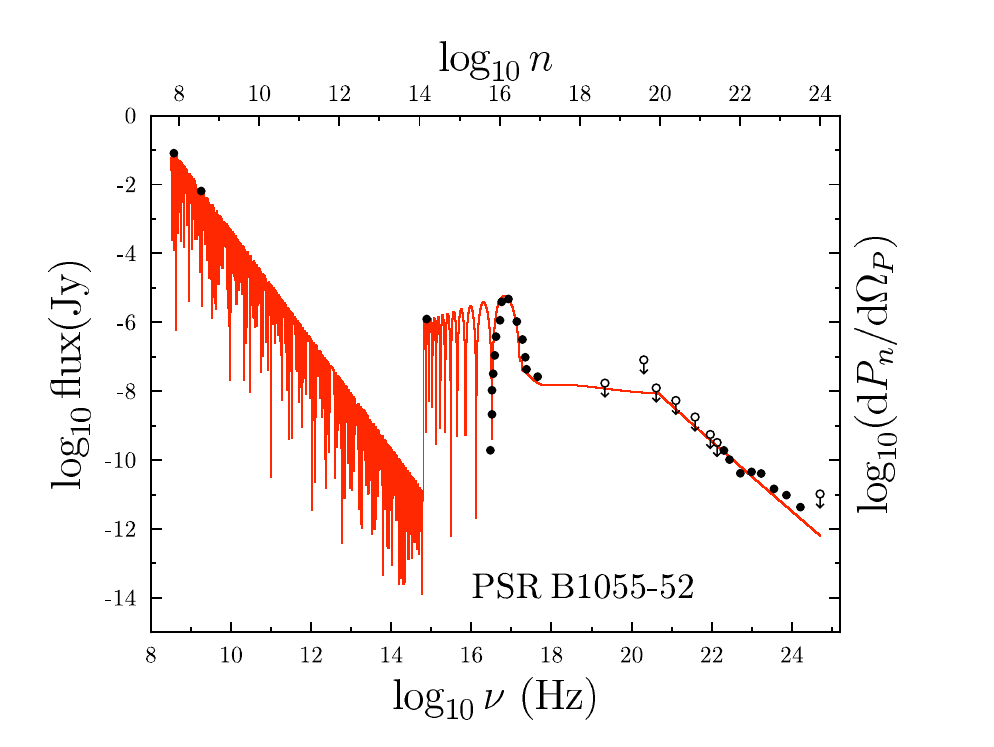}
\caption{The points show observational data (where available) from the spectrum of B1055-52.  The cuves show the spectral distribution $\log({\rm d}P_n/{\rm d}\Omega_P)$, predicted by equations (\ref{eq:16here}), (\ref{eq:33}) and (\ref{eq:34}), versus $\log n$ and $\log \nu$ for $\nu=n\omega/(2\pi)\simeq 5.07 n$ Hz and $\Omega/\omega\simeq6.31\times10^5$.  In the model, the recovery of intensity in the optical band is caused by resonant enhancement due to the azimuthal modulation frequency $m\omega/(2\pi)\simeq6.6\times10^{14}$ Hz.  The steepening of the gradient of the spectrum by $-1$ at $4.5\times10^{20}$~Hz is due to the crossing of the Rayleigh distance.}
\label{fig12}
\end{figure}

We have taken the data points shown in Fig.~\ref{fig12} from
\citet{b21} and~\citet{b37} [see also~\citet{b17}]. 
 The curve in Fig.~\ref{fig12}
represents the radiation flux given by equation (\ref{eq:16here})
for the following values of the parameters: 

\begin{equation}
\frac{\Omega}{\omega}\simeq6.31\times10^5\quad {\rm with}\quad \frac{\omega}{2\pi}=5.07\, {\rm Hz},
\label{eq:32}
\end{equation}

\begin{eqnarray}
S_1\propto\left\{ \begin{array}{lrll}
n^{-3/8}     & 6.3\times10^7 &< n <& 1.3\times10^{14}\\
n^{1/10}      & 1.3\times10^{14}&< n <& 3.7\times 10^{16}\\
n^0     & 3.7\times10^{16}&< n <& 8.9\times 10^{19}\\
n^{-1/2}    & 8.9\times 10^{19} &< n <& 10^{24},
\end{array}\right.
\label{eq:33}
\end{eqnarray}

\begin{eqnarray}
\frac{S_2}{S_1}=\left\{ \begin{array}{lc}
0             & 1.3\times10^{14} < n < 3.7\times 10^{16}\\
4.22\times10^{5}             & \mbox{otherwise,}
\end{array}\right.
\label{eq:34}
\end{eqnarray}
$S_3\ll S_1$ for all $n$, and $m=1.3\times10^{14}$.
 
Given that PSR B1055-52 is at the distance 
$R_P\simeq 4.7\times10^{21}$~cm~\citep{b22} and has a 
light cylinder with the radius 
$c/\omega= 9.4\times10^8$~cm, 
the transition through the Rayleigh distance 
would account for the observed 
steepening of its spectrum at $n=8.9\times10^{19}$ 
(Fig.~\ref{fig12}) if the radial extent ${\hat r}_>-{\hat r}_<$ 
of the emitting plasma is a fraction $4.2\times10^{-4}$ 
of the light-cylinder radius.   

\subsection{PSR B1509-58}
\label{sec:4.7}

\begin{figure}
\centering
\includegraphics[height=6cm]{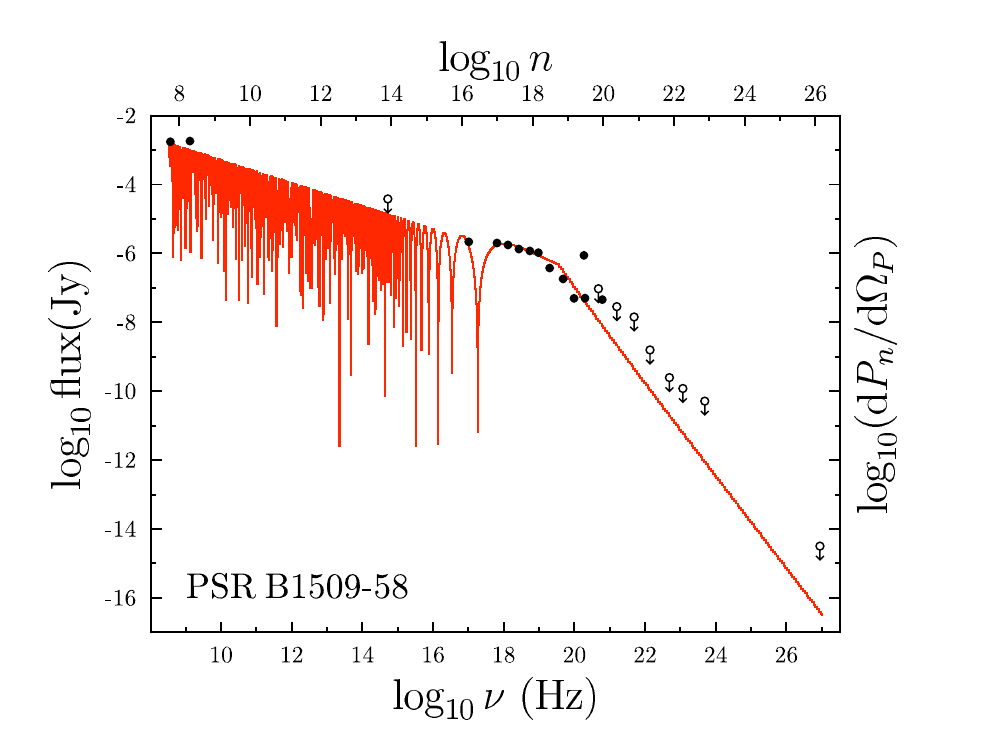}
\caption{The points show observational data (where available) from the spectrum of B1509-58.  The curves show the spectral distribution $\log({\rm d}P_n/{\rm d}\Omega_P)$, predicted by equations (\ref{eq:16here}), (\ref{eq:40}) and (\ref{eq:41}), versus $\log n$ and $\log \nu$ for $\nu=n\omega/(2\pi)\simeq 6.6 n$ Hz and $\Omega/\omega\simeq5.62\times10^5$.  The steepening of the gradient of the spectrum by $-1$ at $3.3\times10^{19}$ Hz is caused by crossing the Rayleigh distance.}
\label{fig13}
\end{figure}

We have taken the data points shown in Fig.~\ref{fig13} from
\citet{b21} [see also~\citet{b17}]. 
The curve in Fig.~\ref{fig13}
represents the radiation flux given by equation (\ref{eq:16here})
for the following values of the parameters: 

\begin{equation}
\frac{\Omega}{\omega}\simeq5.62\times10^5\quad {\rm with}\quad \frac{\omega}{2\pi}=6.6\, {\rm Hz},
\label{eq:39}
\end{equation}

\begin{eqnarray}
S_1\propto\left\{ \begin{array}{lrll}
n^{-1/4}     & 5\times10^7& < n <& 2.5\times10^{15}\\
n^{-1/6}      & 2.5\times10^{15}&< n <& 5\times 10^{18}\\
n^{-2/3}     & 5\times10^{18}&< n <& 1.6\times10^{26},
\end{array}\right.
\label{eq:40}
\end{eqnarray}

\begin{equation}
\frac{S_2}{S_1}\ll1,\quad\frac{S_3}{S_1}\ll1
\label{eq:41}
\end{equation}
for all $n$.  (The existing data is insufficient to pinpoint 
the value of $m$ for this pulsar.)
 
Given that PSR B1509-58 is at the distance 
$R_P\simeq 1.3\times10^{22}$~cm~\citep{b22} and has a 
light cylinder with the radius 
$c/\omega= 7.2\times10^8$~cm, 
the transition through the Rayleigh distance 
would account for the observed 
steepening of its spectrum at $n=5\times10^{18}$ 
(Fig.~\ref{fig13}) if the radial extent ${\hat r}_>-{\hat r}_<$ 
of the emitting plasma is a fraction $3.4\times10^{-3}$ 
of the light-cylinder radius.   

\subsection {PSR B1706-44}
\label{sec:4.4}

\begin{figure}
\centering
\includegraphics[height=6cm]{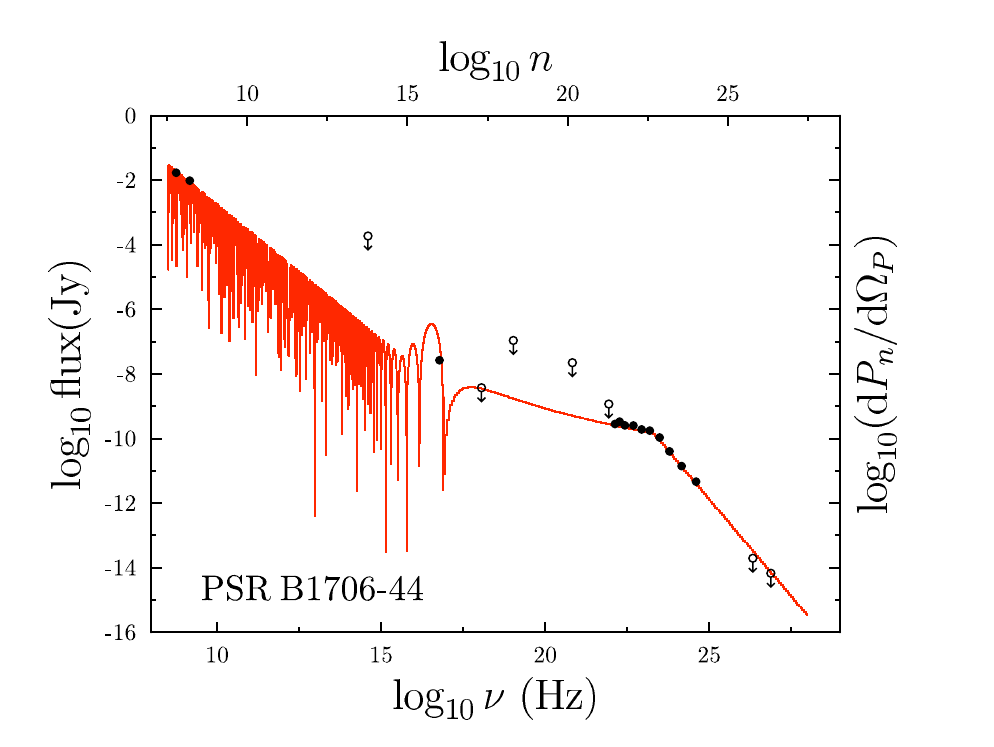}
\caption{The points show observational data (where available) from the spectrum of B1706-44.  The curves show the spectral distribution $\log({\rm d}P_n/{\rm d}\Omega_P)$, predicted by equations (\ref{eq:16here}), (\ref{eq:30}) and (\ref{eq:31}), versus $\log n$ and $\log \nu$ for $\nu=n\omega/(2\pi)\simeq 9.76 n$ Hz and $\Omega/\omega\simeq3.71\times10^5$.  In the model, the recovery of intensity in the optical band is caused by resonant enhancement due to the azimuthal modulation frequency $m\omega/(2\pi)\simeq3.1\times10^{15}$ Hz.  The transit through the Rayleigh distance accounts for the steepening of the gradient of the spectrum by $-1$ at $1.9\times10^{23}$ Hz.}
\label{fig14}
\end{figure}

We have taken the data points shown in Fig.~\ref{fig14} from
\citet{b21} [see also~\citet{b17}]. 
 The curve in Fig.~\ref{fig14}
represents the radiation flux given by equation (\ref{eq:16here})
for the following values of the parameters: 

\begin{equation}
\frac{\Omega}{\omega}\simeq3.71\times10^5\quad {\rm with}\quad \frac{\omega}{2\pi}=9.76\, {\rm Hz},
\label{eq:29}
\end{equation}

\begin{eqnarray}
S_1\propto\left\{ \begin{array}{lrll}
n^{-1/2}     & 3.2\times10^7& < n <& 3.2\times10^{14}\\
n^{1/2}      & 3.2\times10^{14}&< n <& 8.3\times 10^{15}\\
n^{-1/10}     & 8.3\times10^{15}&< n <& 2\times 10^{22}\\
n^{-3/5}    & 2\times 10^{22}& < n <& 10^{27},
\end{array}\right.
\label{eq:30}
\end{eqnarray}

\begin{equation}
\frac{S_2}{S_1}\ll1,\quad\frac{S_3}{S_1}\ll1
\label{eq:31}
\end{equation}
for all $n$, and $m=3.2\times10^{14}$.

Given that PSR B1706-44 is at the distance 
$R_P\simeq 5.6\times10^{21}$~cm~\citep{b22} and has a 
light cylinder with the radius 
$c/\omega= 4.9\times10^8$~cm, 
the transition through the Rayleigh distance 
would account for the observed 
steepening of its spectrum at $n=2\times10^{22}$ 
(Fig.~\ref{fig14}) if the radial extent ${\hat r}_>-{\hat r}_<$ 
of the emitting plasma is a fraction $4.2\times10^{-5}$ 
of the light-cylinder radius.   

\subsection{PSR B1929+10}
\label{sec:4.6}

\begin{figure}
\centering
\includegraphics[height=6cm]{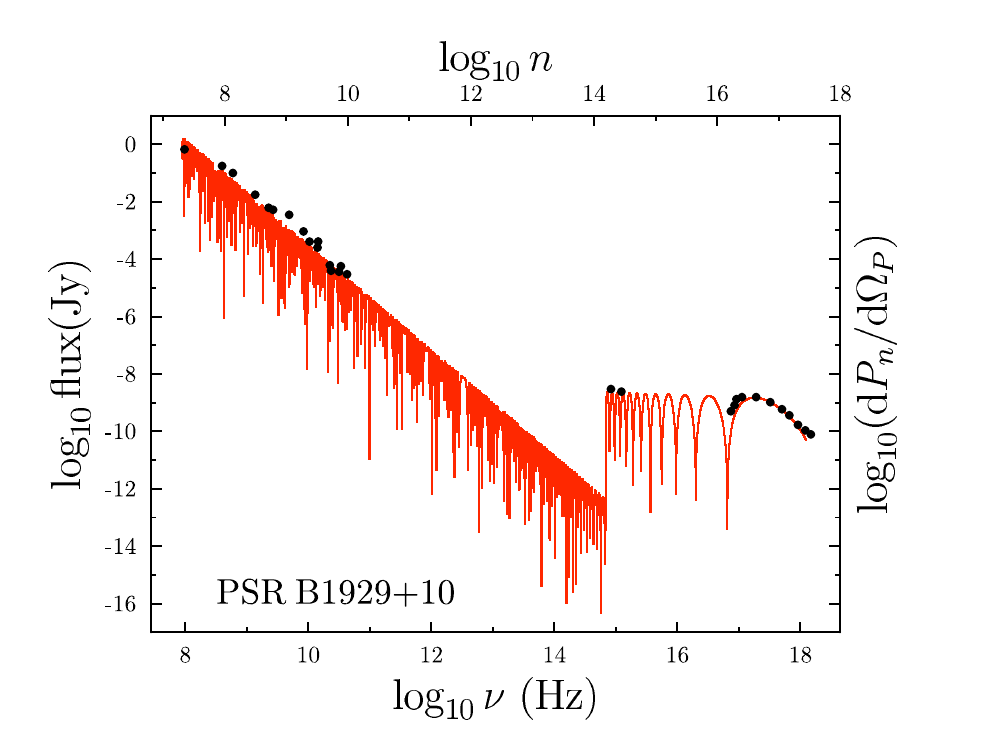}
\caption{The points show observational data (where available) from the spectrum of B1929+10.  The curves show spectral distribution $\log({\rm d}P_n/{\rm d}\Omega_P)$, predicted by equations (\ref{eq:16here}), (\ref{eq:36}) and (\ref{eq:37}), versus $\log n$ and $\log \nu$ for $\nu=n\omega/(2\pi)\simeq 4.41 n$ Hz and $\Omega/\omega\simeq6.3\times10^5$.  In the model, the recovery of intensity in the optical band is caused by resonant enhancement due to the azimuthal modulation frequency $m\omega/(2\pi)\simeq7\times10^{14}$ Hz.}
\label{fig15}
\end{figure}

We have taken the data points shown in Fig.~\ref{fig15} from
\citet{b24}. 
 The curve in Fig.~\ref{fig15}
represents the radiation flux given by equation (\ref{eq:16here})
for the following values of the parameters: 

\begin{equation}
\frac{\Omega}{\omega}\simeq6.3\times10^5\quad {\rm with}\quad \frac{\omega}{2\pi}=4.41\, {\rm Hz},
\label{eq:35}
\end{equation}
\begin{equation}
S_1\propto n^{-1/2},\,\frac{S_2}{S_1}=3.16\times10^6,\quad 2\times10^7 < n < 1.6\times10^{14},
\label{eq:36}
\end{equation}

\begin{equation}
S_2\propto n^{3/8},\,\frac{S_1}{S_2}\ll1,\quad 1.6\times10^{14} < n < 2.8\times10^{17}
\label{eq:37}
\end{equation}
$S_3$ is negligible for all $n$, and $m=1.6\times10^{14}$.

Given that PSR B1929+10 is at the distance 
$R_P\simeq 5.2\times10^{20}$~cm~\citep{b22} and has a 
light cylinder with the radius 
$c/\omega= 1.1\times10^9$~cm, the absence of a sudden steepening
in the gradient of its spectrum (Fig.~\ref{fig15}) 
means that the transition through the Rayleigh distance 
must occur at harmonic numbers larger than 
$n=2.8\times10^{17}$, and so
the radial extent ${\hat r}_>-{\hat r}_<$ 
of the emitting plasma must be smaller than a fraction 
$2.3\times10^{-3}$ of the light-cylinder radius.    

\subsection{PSR B1951+32}
\label{sec:4.9}

\begin{figure}
\centering
\includegraphics[height=6cm]{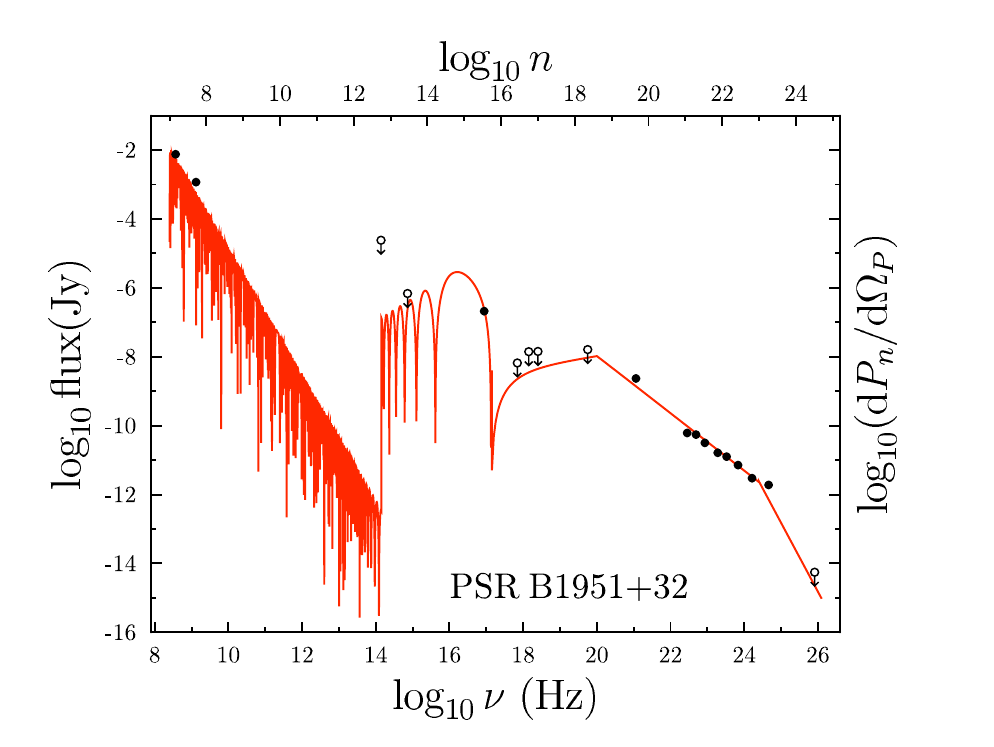}
\caption{The points show observational data (where available) from the spectrum of B1951+32.  The curves show the spectral distribution $\log({\rm d}P_n/{\rm d}\Omega_P)$, predicted by equations (\ref{eq:16here}), (\ref{eq:47}) and (\ref{eq:48}), versus $\log n$ and $\log \nu$ for $\nu=n\omega/(2\pi)\simeq 25.3 n$ Hz and $\Omega/\omega\simeq1.41\times10^5$.  In the model, the recovery of intensity in the optical band is caused by resonant enhancement due to the azimuthal modulation frequency $m\omega/(2\pi)\simeq1.4\times10^{14}$ Hz.  The crossing of the Rayleigh distance accounts for the steepening of the gradient of the spectrum by $-1$ at $10^{20}$ Hz.}
\label{fig16}
\end{figure}

We have taken the data points shown in Fig.~\ref{fig16} from
\citet{b21} [see also~\citet{b17}]. 
The curve in Fig.~\ref{fig16}
represents the radiation flux given by equation (\ref{eq:16here})
for the following values of the parameters: 

\begin{equation}
\frac{\Omega}{\omega}\simeq1.41\times10^5\quad {\rm with}\quad \frac{\omega}{2\pi}=25.3\, {\rm Hz},
\label{eq:46}
\end{equation}

\begin{eqnarray}
S_2\propto\left\{ \begin{array}{lrll}
n^{-1/2}     & 10^7& < n <& 5.6\times10^{12}\\
n^{3/4}      & 5.6\times10^{12}&< n <& 5.6\times 10^{15}\\
n^{5/12}     & 5.6\times10^{15}&< n <& 4\times10^{18}\\
n^{-1/12}     & 4\times10^{18}&< n <& 10^{23}\\
n^{-2/3}     & 10^{23}&< n <& 5\times10^{24},
\end{array}\right.
\label{eq:47}
\end{eqnarray}

\begin{equation}
\frac{S_1}{S_2}\ll1,\quad\frac{S_3}{S_2}\ll1
\label{eq:48}
\end{equation}
for all $n$, and $m=5.6\times10^{12}$.

Given that PSR B01951+32 is at the distance 
$R_P\simeq 7.7\times10^{21}$~cm~\citep{b22} and has a 
light cylinder with the radius 
$c/\omega= 1.9\times10^8$~cm, 
the transit through the Rayleigh distance
would account for the observed 
steepening of its spectrum at $n=4\times10^{18}$ 
(Fig.~\ref{fig16}) if the radial extent ${\hat r}_>-{\hat r}_<$ 
of the emitting plasma is a fraction $5.6\times10^{-3}$ 
of the light-cylinder radius.   

\section{Discussion}
\label{sec:5}
\subsection{Fitted values of $\Omega$ and inferred plasma frequencies}
The fitted values of $\Omega/\omega$
taken from Section~\ref{sec:4} are summarized in
Table~\ref{table3},
where we have also converted the $\Omega/ 2\pi$
values into Hz using the known rotation frequencies of the pulsars.
It will be remembered that $\Omega/ 2 \pi$ represents 
a characteristic frequency 
that modulates the source [see equation~(\ref{eq:3})]
in the emitting region;
assuming that the pulsar
magnetosphere consists of a dilute neutral plasma~\citep{b2},
it is reasonable to make the attribution 
$\Omega/2\pi = f_{\rm p}$, where $f_{\rm p}$
is the (electron) plasma frequency in the emitting part of
the magnetosphere.
The fifth column of Table~\ref{table3}
gives inferred electron densities $N_{\rm e}$ derived using 
$(2\pi f_{\rm p})^2= 4 \pi N_{\rm e}e^2/m_{\rm e}$,
where $e$ is the electron's charge and $m_{\rm e}$ is
its mass~\citep{b17}. Note that virtually all of the electron densities
are of the order of $10^5{\rm cm}^{-3}$,
with only the Crab's value ($N_{\rm e} = 4\times 10^{3} {\rm cm}^{-3}$)
being somewhat lower.\footnote{The lower electron density
around the Crab could result from its very high optical
luminosity~\citep{b17}, which might act to ``blow''
electrons out of the magnetosphere.} Nevertheless, all of these $N_{\rm e}$ are
completely consistent with the densities expected in a conventional
pulsar magnetosphere~\citep{b17}.

\begin{table}
\caption{Summary of $\Omega$ values used in fitting
pulsar spectra.
The parameter $f_{\rm p}$ is the 
derived plasma
frequency, where we assume
that $f_{\rm p}\equiv \Omega/2\pi$.
$N_{\rm e}$ is the electron density
in the emitting region of the pulsar's atmosphere
calculated from the plasma frequency $f_{\rm p}$,
using $(2\pi f_{\rm p})^2= 4 \pi N_{\rm e}e^2/m_{\rm e}$,
where $e$ is the electron's charge and $m_{\rm e}$ is
its mass.}
\centering
\begin{tabular}{lclclclclcl}
\hline
Pulsar & $\frac{\Omega}{\omega}$ & $\frac{\omega}{2 \pi}$ & $f_{\rm p}$ & 
$N_{\rm e} $ \\
 & & (Hz) & (MHz) & ($10^4$cm$^{-3}$)\\
\hline
\hline
PSR B0531+21 & 19000 & 30.3 & 0.58 & 0.4 \\
\hline
PSR B0833-45 & 282000 & 11.2 & 3.2 & 12\\
\hline
PSR J0633+1746 & 589000 & 4.22 & 2.5 & 7.7\\
\hline
PSR B0656+14 & 794000 & 2.6 & 2.1 & 5.3\\
\hline
PSR B1055-52 & 631000 & 5.07 & 3.2 & 13\\
\hline
PSR B1509-58 & 562000 & 6.6 & 3.7 &17\\
\hline
PSR B1706-44 & 371000 & 9.76 & 3.6 & 16\\
\hline
PSR B1929+10 & 630000 & 4.41 & 2.8 & 9.6\\
\hline
PSR B1951+32 & 141000 & 25.3 & 3.6 &16\\
\hline
\label{table3} 
\end{tabular} 
\end{table}

\begin{figure}
\centering
\includegraphics[height=6cm]{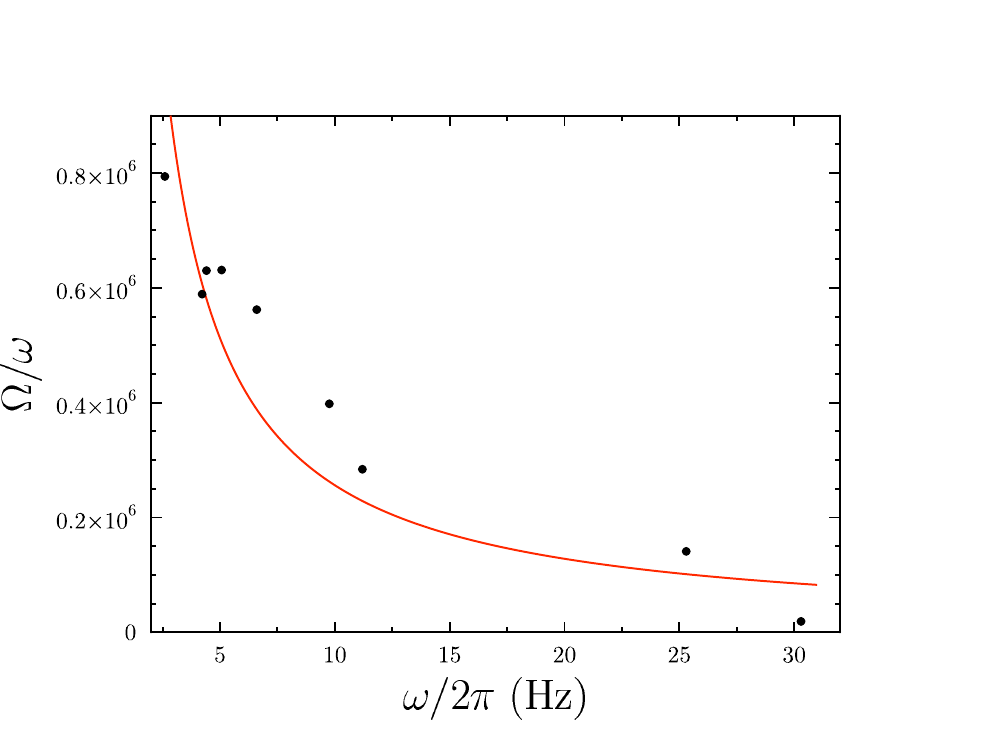}
\caption{The points designate the values of the
two parameters $\Omega/\omega$ and $\omega/2\pi$ for the 9 pulsars 
considered in Section \ref{sec:4}, and the curve 
$\Omega/\omega=2.56\times10^6(2\pi/\omega)$ represents 
the best fit to these points.}
\label{fig17}
\end{figure}

Another way of visualizing the 
rather similar values of $\Omega$ found in Section~\ref{sec:4} for
all of the pulsars is to plot $\Omega/\omega$
versus $\omega/2\pi$ (Fig.~\ref{fig17}); the values of 
$\Omega/\omega$ all lie close to 
the curve $\Omega/\omega=2.56\times10^6(2\pi/\omega)$.
 
Given that positions of the last peaks of the Airy functions appearing in 
equation~(\ref{eq:16here}) and hence the salient features of the 
pulsar spectrum scale as $(\Omega/\omega)^3$, the
rather uniform values of $\Omega$ imply
that slower pulsars (those with smaller values of $\omega$)
will have observed spectral intensities weighted towards higher 
frequencies.  By contrast, millisecond pulsars (with large 
$\omega$) should have emission concentrated at lower frequencies
than the above pulsars.  These predictions seem to be borne out 
both by the radio-quiet pulsar in the supernova remnant CTA1
($\omega/2\pi\simeq 3.1$ Hz), which has emission 
peaked in the gamma-ray end of the spectrum~\citep{b38}, and by 
millisecond pulsars such as J1748-22446ad ($\omega/2\pi\simeq716$
Hz), B1937+21 ($\omega/2\pi\simeq642$
Hz) and B1957+20 ($\omega/2\pi\simeq622$ Hz) which show no 
emission in the high-frequency range but generally strong 
and bright pulses at radio frequencies~\citep{b39,b40,b41}. 
\subsection{The frequency $m\omega/2\pi$ and inferred magnetic fields}
Table~\ref{table4} shows
the values of $m$ derived from the frequencies at which 
the observed spectra recover their intensity
as a result of resonance enhancement
(see Section~\ref{sec:4}).
The frequency $m\omega/2\pi$ represents the frequency of azimuthal
fluctuations of the polarization-current distribution pattern.
These spatial fluctuation frequencies 
($\sim10^{13}-10^{15}$~Hz) could well arise from
the cyclotron resonance of electrons. The cyclotron
(angular) frequency $\omega_{\rm c}$ 
in a field $H$ is defined as
$\omega_{\rm c} =eH/cm_{\rm e}$, where $m_{\rm e}$
is the mass of the electron and $e$ is its charge (cgs units)~\citep{b17};
hence, we can use the $m\omega$ values to infer
magnetic fields of around $10^7-10^9$~G in the 
emitting region~(Table~\ref{table4}). 
Such fields are
completely consistent with those expected in a conventional
pulsar magnetosphere~\citep{b17}.
\begin{table}
\caption{Summary of $m$ values used in fitting
pulsar spectra.
The final column tabulates
the magnetic fields deduced on the assumption that
$m \omega/2 \pi$ represents the cyclotron resonance 
of electrons
in the emitting region of the magnetosphere.
These were derived using
the standard equation for the cyclotron (angular)
frequency $\omega_{\rm c}$:
$m\omega = \omega_{\rm c} \equiv eH/cm_{\rm e}$,
where $e$ is the electron's charge and $m_{\rm e}$ is
its mass.}
\centering
\begin{tabular}{lclclclclcl}
\hline
Pulsar & $m$ & $\frac{\omega}{2 \pi}$ & $\frac{m\omega}{2\pi}$ & 
$H$ \\
 & & (Hz) & (THz) & (G)\\
\hline
\hline
PSR B0531+21 & $10^{12}$ & 30.3 & 30 & $1.1\times 10^7$ \\
\hline
PSR B0833-45 & $1.3\times 10^{15}$ & 11.2 & 15000 & $5.2\times10^9$\\
\hline
PSR J0633+1746 & $1.6\times 10^{14}$ & 4.22 & 680 & $2.4\times 10^8$\\
\hline
PSR B0656+14 & $2\times 10^{14}$ & 2.6 & 520 & $1.9\times 10^8$\\
\hline
PSR B1055-52 & $1.3\times 10^{14}$ & 5.07 & 660 & $2.4\times10^8$\\
\hline
PSR B1509-58 & -- & 6.6 & -- &--\\
\hline
PSR B1706-44 & $3.2\times 10^{14}$ & 9.76 & 3100 & $1.1\times 10^9$\\
\hline
PSR B1929+10 & $1.6\times 10^{14}$ & 4.41 & 710 & $2.5\times 10^8$\\
\hline
PSR B1951+32 & $5.6\times 10^{12}$ & 25.3 & 140 & $5.1\times 10^7$\\
\hline
\label{table4} 
\end{tabular} 
\end{table}
\subsection{The size of the emitting region}
The set of frequencies at which the slopes of the observed spectra 
suddently steepen by $-1$ mostly lie in the range 
$10^{18}-10^{21}$ Hz.  This implies that the radial extent
of the emitting polarization current is a 
fraction $10^{-2}-10^{-5}$ of the light-cylinder radius 
(Section~\ref{sec:4}). Recalling that the subbeam that arises
from the emitting part of the source in the radial interval
${\hat r}_<\leq{\hat r} \leq{\hat r}_>$ (where ${\hat r}\equiv r\omega/c$) 
is detectable within the polar interval
$\arccos(1/{\hat r}_<)\le\vert\theta_P-\pi/2\vert
\le\arccos(1/{\hat r}_>)$ (see Section~\ref{sec:2.3.1}),  
it follows that the polar widths
$\Delta\theta_P\simeq\sin\theta_P\tan\theta_P({\hat r}_>-{\hat r}_<)$ 
of the subbeams that are detected  
outside the plane of rotation ($\theta_P\neq\pi/2$)
are of the order of $10^{-2}-10^{-5}$ radians.
[Note that the overall radiation beam consists of a superposition
of such narrow subbeams (Section~\ref{sec:2.3.1}).] 
\subsection{Frequency dependence of Fourier components of the source density ${\bf s}$}
\label{freqdep}
The dependence of the Fourier components ${\bar s}_{r,\varphi,z}$ 
of the source densities $s_{r,\varphi,z}$ on frequency 
(or on harmonic number $n$) is determined by the fluctuations 
(within the emitting region) of $s_{r,\varphi,z}$ with the 
coordinate $z$ [see equation (\ref{eq:a25})].
The small-scale spatial variations 
of ${\bf s}$ in the direction parallel to the rotation 
axis that are implied by the observational data 
(see Section~\ref{sec:4}) may be identified 
with the sharp gradients in the distribution of
the source density at a current sheet.  The two-dimensional
current sheets predicted by 
numerical models of MHD pulsar magnetospheres
extend beyond the light cylinder.
Not only do they have distribution patterns that rotate 
superluminally~\citep{b2}, 
but they also vary discontinuously, and so have 
fluctuations that occur over 
wide ranges of length scales and amplitudes.

\section{Summary and conclusion}
\label{sec:6}

We have compared the observed spectra of 9 pulsars for which 
multiwavelength data are available with the spectrum of the 
radiation generated by an extended source in superluminal rotation 
and shown that the entire spectrum of each pulsar can be accounted for 
quantitatively in terms of this single emission process. 
The results reported here are model-independent in that the only 
global property of the magnetospheric structure invoked is its  
quasi-steady time dependence: the cylindrical components 
$j_{r,\varphi,z}(r,\varphi,z;t)$ of the density of the magnetospheric 
polarization current depend on $\varphi$ only in the combination 
$\varphi-\omega t$. This property follows unambiguously from 
the observational data and implies that a current distribution with a 
superluminally rotating pattern at a radius $r > c/\omega$ is responsible 
for the unique features of pulsar emission such as the pulses' extreme 
brightness temperature, temporal width, source dimension and 
peak spectral frequency as well as the average pulses' polarization properties 
(their occurrence as concurrent 'orthogonal' modes with swinging position 
angles and with nearly 100 per cent linear or circular polarization).

The curves in Figs.~\ref{fig8}--\ref{fig16} demonstrate 
that the universal features of pulsar spectra (their breadth over  
15--18 orders of magnitude of frequency, the recovery of their 
amplitudes in the optical band,
their bell-shaped peaks in ultraviolet, and the sudden 
steepening of their gradient 
by $-1$ in the $X$-ray band) are all described by the properties 
of the Green function for the problem, and 
so are consequences solely of the accelerated
superluminal motion of the distribution pattern of their source. Only 
the finer details of these spectra depend on the structure of the
pulsar magnetosphere, or more specifically, on the structure of the current 
sheets just outside the light cylinder of the pulsar magnetosphere~\citep{b2} 
(i.e. on the variable `weather' in the pulsar magnetosphere, 
as opposed to its stable 
`climate').  

The increase in the spacing between successive peaks of the 
oscillating spectra shown in Figs.~\ref{fig8}--\ref{fig16} with frequency,
which has already been observed in the radio emission
from the Crab pulsar~\citep{b16}, also
only depends on the properties of the Airy functions in equation
(\ref{eq:16here}); it is another universal feature of the pulsar 
emission predicted by the present model that can be tested, in principle,
over any frequency band [see Fig.~\ref{fig7} and~\citet{b1}].

\section{Acknowledgments}

We are grateful to Quinn Marksteiner for helpful comments.  
This work is supported by U.S. Department of 
Energy Grant LDRD 20080085DR,
``Construction and use of superluminal emission 
technology demonstrators
with applications in radar, 
astrophysics and secure communications''.
A.\ Ardavan thanks the Royal Society for support.

\appendix
\section{ Mathematical description of the radiated field}

A general mathematical treatment of spectrum of the radiation 
field that is generated by the source distribution described 
in equation (\ref{eq:3}) has already been given in~\citet{b10}.  
Our purpose in this appendix is to make the physical content of 
the previous analysis more transparent by presenting its results
from an alternative point of view.  We begin by calculating the 
spectrum of the radiation emitted by one 
of the constituent rings of source (\ref{eq:3}): 
by first deriving the 
spectrum of the Li\'enard-Wiechert potential due to
a superluminally rotating point source, and then 
superposing the potentials of the point-like line 
elements that make up the ring.
Once we have clarified the origin of the higher 
frequencies in the emission from this simpler ring
current, we will further superpose the 
fields of the rings that make up the volume-distributed 
source (\ref{eq:3}) to arrive at the spectrum on which 
the modelling of the observed spectra considered 
in this paper is based. 

\subsection{Radiation field of a constituent ring of the source}
\label{sec:A1}

The trajectory of a uniformly rotating point source, 
e.g.\ the source of synchrotron radiation, is described by
\begin{equation}
r=r_0,\quad z=z_0,\quad\varphi=\varphi_0+\omega t,
\label{eq:a1}
\end{equation}
where the subscript $0$ denotes the constant values of the 
cylindrical polar coordinates $(r,\varphi,z)$ of the point 
source at time $t=0$, and $\omega$ is constant. (Note that, 
to describe repeated rotations, $\varphi$ in this equation 
increases indefinitely as the time $t$ does.) If this 
source has a strength $s$ that oscillates with the frequency
$\Omega$ in time, then the density of the polarization current 
it carries can be written as
\begin{equation}
{\bf j}=s\omega\exp({\rm i}\Omega t)\delta(r-r_0)\delta(z-z_0)
\delta(\varphi-\varphi_0-\omega t){\hat{\bf e}}_\varphi,
\label{eq:a2}
\end{equation}
where $\delta$ is the Dirac delta function, and ${\hat{\bf e}}_\varphi$ is the base vector associated with the azimuthal coordinate $\varphi$. 

The vector potential due to this oscillating, rotating 
point source is given by
\begin{eqnarray}
{\bf A}^{\rm point}&=&\frac{1}{c}\int r{\rm d}r{\rm d}\varphi{\rm d}z{\rm d}t\,{\bf j}\,\frac{\delta(t_P-t-R/c)}{R}\nonumber\\
&=&\frac{sr_0\omega}{c}\int{\rm d}t\exp({\rm i}\Omega t)\frac{\delta(t_P-t-R_0/c)}{R_0}{\hat{\bf e}}_{\varphi0},
\label{eq:a3}
\end{eqnarray}
in which
\begin{equation}
R_0=R\vert_{r=r_0,\varphi=\varphi_0+\omega t,z=z_0},
\,{\hat{\bf e}}_{\varphi0}={\hat{\bf e}}_{\varphi}
\vert_{r=r_0,\varphi=\varphi_0+\omega t,z=z_0},
\label{eq:a4}
\end{equation}
with
\begin{equation}
R=[(z_P-z)^2+{r_P}^2+{r}^2-2rr_P\cos(\varphi_P-\varphi)]^{1/2},
\label{eq:a5}
\end{equation}
and
\begin{equation}
{\hat{\bf e}}_\varphi=\sin(\varphi_P-\varphi){\hat{\bf e}}_{r_P}
+\cos(\varphi_P-\varphi){\hat{\bf e}}_{\varphi_P},
\label{eq:a6}
\end{equation}
where $({\hat{\bf e}}_{r_P},{\hat{\bf e}}_{\varphi_P},{\hat{\bf e}}_{z_P})$ are 
the constant base vectors associated with the coordinates 
$(r_P,\varphi_P,z_P)$ at the observation point $P$, and 
$c$ is the speed of light {\it in vacuo}.  Evaluation 
of the remaining integral with respect to $t$ yields
\begin{eqnarray}
{\bf A}^{\rm point}&=&\frac{sr_0\omega}{c}\sum_{t=t_j}\frac{\exp({\rm i}\Omega t)}{\vert R_0-(r_0r_P\omega/c)\sin(\varphi_P-\varphi_0-\omega t)\vert}\nonumber\\
&&\times[\sin(\varphi_P-\varphi_0-\omega t){\hat{\bf e}}_{r_P}\nonumber\\
&&+\cos(\varphi_P-\varphi_0-\omega t){\hat{\bf e}}_{\varphi_P}],
\label{eq:a7}
\end{eqnarray}
where the retarded times $t_j$ are the solutions of
\begin{equation}
t_P=t+R_0/c
\label{eq:a8}
\end{equation}
[see, e.g.~\citet{b18}].  Depending on the values of $(r_0,\varphi_0,z_0)$, 
there are either one or three $t_j$s when the speed $r_0\omega$ 
of the source exceeds $c$ moderately [see Fig.~\ref{fig1}, and \citet{b9}].

Taking the curl of ${\bf A}^{\rm point}$, and noting 
that the dependence of this potential on the 
coordinates of a far-field observer 
arises primarily from the dependence of the retarded times 
$t_j$ on these coordinates, we obtain the following expression 
for the magnetic field of the radiation emitted by the above 
point source:
\begin{eqnarray}
{\bf B}^{\rm point}&=&-\frac{sr_0\omega^2}{c^3}\sum_{t=t_j}\frac{1}{\vert 1-{\hat r}_0{\hat r}_P\sin(\varphi_P-\varphi_0-\omega t)/{\hat R}_0\vert}\nonumber\\
&&\times\frac{{\rm d}}{{\rm d}t}\left[\frac{\exp({\rm i}\Omega t){\bf p}}{{\hat R}_0-{\hat r}_0{\hat r}_P\sin(\varphi_P-\varphi_0-\omega t)}\right],
\label{eq:a9}
\end{eqnarray}
with 
\begin{equation}
{\bf p}=\cos\theta_P\sin(\varphi_P-\varphi_0-\omega t){\hat{\bf e}}_\parallel+\cos(\varphi_P-\varphi_0-\omega t){\hat{\bf e}}_\perp,
\label{eq:a9b}
\end{equation}
where ${\hat{\bf e}}_\parallel={\hat{\bf e}}_{\varphi_P}$ (which is parallel to the plane of rotation) and ${\hat{\bf e}}_\perp={\hat{\bf n}}{\bf\times}{\hat{\bf e}}_\parallel$ comprise a pair of unit vectors normal to the radiation direction
\begin{equation}
{\hat{\bf n}}=\sin\theta_P{\hat{\bf e}}_{r_P}+\cos\theta_P{\hat{\bf e}}_{z_P},
\label{eq:a10}
\end{equation}
and ${\hat r}_0=r_0\omega/c$, ${\hat r}_P=r_P\omega/c$, and ${\hat R}_0=R_0\omega/c$.  This is the familiar Li\'enard-Wiechert field, except that it receives contributions from more than one retarded time $t_j$ if ${\hat r}_0>1$ and so the source moves faster than light (see Section~\ref{sec:2}).

Next, let us consider a rotating ring $r=r_0$, $z=z_0$, whose 
strength varies spatially, as $\cos(m\varphi_0)$, 
in addition to oscillating temporally  
($m$ is an integer).  The field due to such a ring 
can be obtained from the field ${\bf B}^{\rm point}$ of the 
above point source by multiplying ${\bf B}^{\rm point}$ by 
$\cos(m\varphi_0)$ and superposing the contributions of the 
elements that make up the ring: elements whose azimuthal 
positions $\varphi_0$ at $t=0$ cover a full circle only 
once, and so are labelled uniquely by their initial coordinates 
over the interval $-\pi<\varphi_0\le\pi$.  That is to say,
\begin{equation}
{\bf B}^{\rm ring}=\int_{-\pi}^\pi{\bf B}^{\rm point}
\cos(m\varphi_0){\rm d}\varphi_0,
\label{eq:a11}
\end{equation}
with a range of integration that is limited to $2\pi$.  
[Note that the variable $\varphi_0$ is the same as the 
variable ${\hat\varphi}$ appearing in equation (\ref{eq:3}).]

For ${\hat r_0}<1$, the denominators in equation (\ref{eq:a9}) 
nowhere vanish, and so ${\bf B}^{\rm point}$ is a regular 
and periodic function of $\varphi_0+\omega t_P$: according to 
equation (\ref{eq:a8}), $t_j-t_P$ are periodic 
functions of $\varphi_0+\omega t_P$ with the period $2\pi/\omega$.  
In this subluminal case, the expression on the right-hand 
side of equation (\ref{eq:a11}) constitutes the coefficient of a single
term in the cosine series for ${\bf B}^{\rm point}$.  Thus 
the Fourier expansion of ${\bf B}^{\rm ring}$ consists of 
only two sinusiodal functions of $\varphi_0+\omega t_P$
with the frequencies $\Omega\pm m\omega$.  For ${\hat r}_0>1$, 
on the other hand, the denominators in equation (\ref{eq:a9}) 
vanish for a source point $(r_0,\varphi_0,z_0)$ that 
approaches the observer with the speed of light and zero 
acceleration at the retarded time: the expression appearing in the 
first denominator in this equation 
equals $1+c^{-1}{\rm d}R_0/{\rm d}t$.  
Being non-integrable~\citep{b11}, this singularity
makes it impossible to Fourier analyze the 
representation (\ref{eq:a9}) of the Li\'enard-Wiechert 
field ${\bf B}^{\rm point}$ directly to obtain the 
spectral distribution of this field.

The way to handle the singularity of ${\bf B}^{\rm point}$ 
in the superluminal case is to go 
back one step and instead expand the integrand in equation 
(\ref{eq:a3}) in a Fourier series with respect to $\varphi_0$.  
(We must use a Fourier series, rather than a Fourier integral, 
because the values of $\varphi_0$ over which this function is 
defined lie in an interval of finite length.)  The result is
\begin{eqnarray}
{\bf A}^{\rm point}&=&\frac{sr_0\omega^2}{2\pi c}\sum_{n=-\infty}^\infty
\exp({\rm i}n\omega t_P)\int{\rm d}t \,{R_0}^{-1}\nonumber\\
&&\times\exp\{-{\rm i}
[n\omega(t+R_0/c)-\Omega t]\}{\hat{\bf e}}_{\varphi_0},
\label{eq:a12}
\end{eqnarray}
as can be readily seen by replacing the delta function in 
equation (\ref{eq:a3}) by its Fourier-series representation.
Taking the curl of this, we then obtain
\begin{eqnarray}
{\bf B}^{\rm point}&=&\frac{sr_0\omega^3}{2\pi{\rm i}c^2}\sum_{n=-\infty}^\infty
n\exp({\rm i}n\omega t_P)\int{\rm d}t \,{R_0}^{-1}\nonumber\\
&&\times\exp\{-{\rm i}[n\omega(t+R_0/c)-\Omega t]\}\nonumber\\
&&\times[\cos\theta_P\sin(\varphi_P-\varphi_0-\omega t){\hat{\bf e}}_\parallel\nonumber\\
&&+\cos(\varphi_P-\varphi_0-\omega t){\hat{\bf e}}_\perp].
\label{eq:a13}
\end{eqnarray}
For a radiation frequency $n\omega/2\pi$ that appreciably 
exceeds the rotation frequency, 
the asymptotic value of the integral in the above expression
receives contributions solely from the stationary points of 
the phase, $ct+R_0(t)$, of the rapidly oscillating exponential
in its integrand.

This phase is stationary, in the course of each rotation, 
at the following two retarded times 
at which the source approaches 
the observer along the radiation direction with the speed $c$: 
\begin{equation}
\omega t_\pm=\varphi_P-\varphi_0+\frac{3\pi}{2}\pm\arccos
\left(\frac{1}{{\hat r}_0\sin\theta_P}\right)+2k\pi,
\label{eq:a14}
\end{equation}
where $(R_P,\theta_P,\varphi_P)$ (with $R_P\gg c/\omega$) denote 
the spherical polar coordinates of the observation point
and $k$ is an integer.
These stationary points coalesce if the observer 
is located at $\theta_P=\arcsin(1/{\hat r}_0)$,
i.e.\ when the source approaches the observer not 
only with the speed $c$ but also with zero 
acceleration [see Fig.~\ref{fig1}, and~\citet{b12}].  
In this case, the 
rapid oscillations of the exponential in equation 
(\ref{eq:a13}) result in the destructive interference of 
contributions from all emission times, except
those made during a short interval centred at $t=t_c$, 
where $\omega t_c=\varphi_P-\varphi_0+3\pi/2+2k\pi$.

For $n\gg1$ and $\theta_P=\arcsin(1/{\hat r}_0)$,
we can therefore obtain the leading term in the asymptotic
expansion of the integral in equation (\ref{eq:a13})
by applying the principle of stationary phase: by approximating 
both the phase $ct+R_0(t)$ and the amplitude of the rapidly 
oscillating exponential with the dominant terms in their 
Taylor expansions about $t=t_c$ and
replacing the limits of integration with $\pm\infty$~\citep{b50}.  
Evaluation of the resulting integral with the aid of 
equation (10.4.32) of~\citet{b19} thus yields
\begin{eqnarray}
{\bf B}^{\rm point}&\simeq&\frac{2sr_0\omega^2}{{\rm i}c^2R_P}\sum_{\vert n\vert\gg1}\left(\frac{n}{2}\right)^{2/3}\nonumber\\
&&\exp\{{\rm i}[n(\omega t_P-{\hat R}_P+{\hat z}_0\cos\theta_P)\nonumber\\
&&+(n-\Omega/\omega)(\varphi_0-\varphi_P-3\pi/2)]\}\nonumber\\
&&\times\left\{\cos\theta_P{\rm Ai}\left[-\left(\frac{2}{n}\right)^{1/3}\frac{\Omega}{\omega}\right]{\hat{\bf e}}_\parallel\right.\nonumber\\
&&\left. +{\rm i}\left(\frac{2}{n}\right)^{1/3}{\rm Ai}^\prime\left[-\left(\frac{2}{n}\right)^{1/3}\frac{\Omega}{\omega}\right]{\hat{\bf e}}_\perp\right\},
\label{eq:a15}
\end{eqnarray}
where Ai and Ai$^\prime$ stand for the Airy function and the derivative of the Airy function with respect to its argument, respectively, and ${\hat R}_P=R_P\omega/c$ and ${\hat z}_0=z_0\omega/c$.

Amplitudes of the individual terms in equation (\ref{eq:a15})
have the dependence $n^{-1/4}$ on $n$ for $n\gg1$, so that 
this series yields a divergent value for ${\bf B}^{\rm point}$ 
on the cusp locus $\theta_P=\arcsin(1/{\hat r}_0)$, as did 
equation (\ref{eq:a13}).  However, these individual Fourier 
components are not periodic functions of $\varphi_0$ when 
$\Omega$ differs from an integral multiple of $\omega$.
As a result, when we insert equation (\ref{eq:a15}) in equation
(\ref{eq:a11}) and carry out the integration with respect to
$\varphi_0$, we obtain an expression that contains all values of ($n\gg1)$:
\begin{eqnarray}
{\bf B}^{\rm ring}&\simeq&\frac{2sr_0\omega^2}{{\rm i}c^2R_P}\sum_{\vert n\vert\gg1}\left(\frac{n}{2}\right)^{2/3}C_{\varphi_0}\nonumber\\
&&\exp\{{\rm i}[n(\omega t_P
-{\hat R}_P+{\hat z}_0\cos\theta_P)\nonumber\\
&&-(n-\Omega/\omega)(\varphi_P+3\pi/2)]\}\nonumber\\
&&\times\left\{\cos\theta_P{\rm Ai}\left[-\left(\frac{2}{n}\right)^{1/3}\frac{\Omega}{\omega}\right]{\hat{\bf e}}_\parallel\right.\nonumber\\
&&\left. +{\rm i}\left(\frac{2}{n}\right)^{1/3}{\rm Ai}^\prime\left[-\left(\frac{2}{n}\right)^{1/3}\frac{\Omega}{\omega}\right]{\hat{\bf e}}_\perp\right\},
\label{eq:a16}
\end{eqnarray}
where
\begin{eqnarray}
C_{\varphi_0}&=&\frac{\sin[(n+m-\Omega/\omega)\pi]}{n+m-\Omega/\omega}\nonumber\\
&&+\frac{\sin[(n-m-\Omega/\omega)\pi]}{n-m-\Omega/\omega}
\label{eq:a17}
\end{eqnarray}
is the factor resulting from the integration over $\varphi_0$.
This series has an infinite number of non-zero terms because, 
on the one hand, the variable $\varphi_0$ that
labels the elements of the ring can only range 
over an interval of length $2\pi$, and on the other hand, the function 
${\bf B}^{\rm point}$ that we are Fourier analyzing has 
different values at opposite ends of such an interval
[see Section 2 of~\citet{b10}]. 

When $\Omega/\omega$ is an integer, only the two terms 
$n=\pm m+\Omega/\omega$ of the series in 
equation (\ref{eq:a16}) are non-zero: the numerators and 
denominators of the two fractions in equation (\ref{eq:a17}) 
simultaneously vanish approaching the limits $\pi$.  But when 
$\Omega/\omega$ is nonintegral, this series diverges,
as expected from the fact that the singularity of the 
time-domain expression for ${\bf B}^{\rm point}$ in
equation (\ref{eq:a9}) is non-integrable.  
There is a radical difference between the series
in equation (\ref{eq:a16}) and its subluminal counterpart.
Not only is it not possible to 
derive the frequency-domain expression for ${\bf B}^{\rm ring}$ by 
directly Fourier analyzing the time-domain expression for this field
[which is divergent when the distance between the observer and the source 
decreases with the speed ${\rm d}R_0/{\rm d}t=-c$], 
but, correspondingly, the transformation $t\to t+R_0(t)/c$ that takes one 
from the frequency-domain expression to the time-domain expression has a 
vanishing Jacobian and so is mathematically impermissible in the 
superluminal case. a25

On the one hand, each source element $\varphi_0$ of the 
ring makes its contribution towards the observed 
field at the discrete set of retarded times 
$\omega t=\varphi_P-\varphi_0+3\pi/2+2k\pi$ 
at which it approaches the observer with the speed of light 
and zero acceleration, i.e.\ {\em periodically} with the 
period $2\pi/\omega$.  On the other hand, temporal modulations of 
the density of the source occur on a time scale $2\pi/\Omega$ 
that is incommensurable with the period $2\pi/\omega$.  
Because the emission time of the element labelled by $\varphi_0$ 
is fixed (via $\omega t=\varphi_P-\varphi_0+3\pi/2+2k\pi$) 
by its initial azimuthal position ($\varphi_0$), the temporal 
[$\exp({\rm i}\Omega t)$] and spatial [$\cos(m\varphi_0)$] 
modulations of this source effectively combine 
into a single variation [$\propto\exp({\rm i}\Omega\varphi_0/\omega)
\cos(m\varphi_0)$].  It is the Fourier decomposition of 
this variation with $\varphi_0$, or equivalently with
the retarded time $t$, that together with the incommensurablity 
of the two periods ($2\pi/\omega$ and $2\pi/\Omega$), and 
the finiteness of the domain of definition of $\varphi_0$, 
results in a spectrum containing all frequencies 
[see Fig.\ 5 of~\citet{b10}].

\subsection{Radiation field of the entire volume of the source}
\label{sec:A2}

In this Subsection, we use a similar procedure to 
find the field generated by the 
volume-distributed source described in equation 
(\ref{eq:3}) for which
\begin{eqnarray}
{\bf j}&=&\partial{\bf P}/\partial t=\textstyle{1\over4}{\rm i}{\bf s}(r,z)
\sum_{\pm}(\Omega\pm m\omega)\exp[{\rm i}(\Omega t\mp m\varphi_0)]\nonumber\\
&&+\{m\to-m, \Omega\to-\Omega\},
\label{eq:a172}
\end{eqnarray}
where the symbol $\{m\to -m,\Omega\to-\Omega\}$ 
designates a term exactly like the one preceding 
it but in which $m$ and $\Omega$ are everywhere 
replaced by $-m$ and $-\Omega$, respectively.
Here, we have replaced ${\hat\varphi}$ by $\varphi_0$ to 
bring out the connection with the ring source considered in 
Subsection~\ref{sec:A1}.

Taking the curl of the first member of equation 
(\ref{eq:a3}) and discarding terms of order $R^{-2}$,
we obtain the following generalization of equation 
(\ref{eq:a13}):
\begin{eqnarray}
{\bf B}&\simeq&\frac{1}{c^2}\int r{\rm d}r{\rm d}\varphi
{\rm d}z{\rm d}t\frac{\delta^\prime(t-t_P+R/c)}{R}
{\hat{\bf n}}{\bf\times j}\nonumber\\
&=&-\frac{{\rm i}\omega^2}{2\pi c^2}\sum_{-\infty}^\infty
n\exp({\rm i}n\omega t_P)\int r{\rm d}r{\rm d}\varphi_0
{\rm d}z{\rm d}t {R_0}^{-1}\nonumber\\
&&\times\exp[-{\rm i}n(\omega t+{\hat R}_0)]
{\hat{\bf n}}{\bf\times j},
\label{eq:a18}
\end{eqnarray}
where
\begin{equation}
R_0=[(z_P-z)^2+{r_P}^2+{r}^2-2rr_P\cos(\varphi_P-\varphi_0-\omega t)]^{1/2},
\label{eq:a19}
\end{equation}
as before.  Note that the change of integration variable from 
$\varphi$ to $\varphi_0=\varphi-\omega t$ in the second member 
of this equation, that is made to emphasize the correspondence 
between equations (\ref{eq:a13}) and (\ref{eq:a18}), has a non-zero 
Jacobian and so is perfectly permissible.

For $n\gg1$ and a far-field observer outside the plane 
of rotation, i.e.\ $\theta_P\neq\pi/2$, the
phase $t+R_0(t)/c$ of the rapidly oscillating exponential in
equation (\ref{eq:a18}) is stationary at $t=t_c$, $r=\csc\theta_P$,
simultaneously, where $\omega t_c=\varphi_P-\varphi_0+3\pi/2+2k\pi$ 
as in Subsection~\ref{sec:A1}. 
Performing the integration with respect to these two variables
by the method of stationary phase, we obtain the following expression
for the electric field ${\bf E}=-{\hat{\bf n}}{\bf\times B}$ 
of the radiation:
\begin{equation}
{\bf E}=\Re\left\{{\tilde{\bf E}}_0+2\sum_{n=1}^\infty
{\tilde{\bf E}}_n\exp(-{\rm i}n{\hat\varphi}_P)\right\},
\label{eq:a20}
\end{equation}
in which
\begin{eqnarray}
{\tilde{\bf E}}_n&\simeq&\frac{m-\Omega/\omega}{2{\hat r}_P}\left(\frac{n}{2}\right)^{2/3}\exp\left\{{\rm i}
[n({\hat R}_P+3\pi/2) \right.\nonumber\\
&& \left.-(\Omega/\omega)(\varphi_P+3\pi/2)]\right\}K_r
{\bf V}\vert_{{\hat r}=\csc\theta_P}\nonumber\\
&&\times\int_{-\pi}^\pi{\rm d}\varphi_0\exp[{\rm i}
(n-m+\Omega/\omega)\varphi_0]\nonumber\\
&&+\{m\to-m,\Omega\to-\Omega\},
\label{eq:a21}
\end{eqnarray}
with
\begin{equation}
K_r\simeq{\hat r}_>-{\hat r}_<,\quad n\ll\pi{\hat R}_P/
({\hat r}_>-{\hat r}_<)^2,
\label{eq:a22}
\end{equation}
or
\begin{equation}
K_r\simeq(2\pi{\hat R}_P/n)^{1/2}\exp(-{\rm i}\pi/4),\, 
n\gg\pi{\hat R}_P/({\hat r}_>-{\hat r}_<)^2,
\label{eq:a23}
\end{equation}
\begin{eqnarray}
{\bf V}&=&\left[{\bar s}_r{\hat{\bf e}}_\parallel
+({\bar s}_\varphi\cos\theta_P-{\bar s}_z\sin\theta_P)
{\hat{\bf e}}_\perp\right]\nonumber\\
&&\times{\rm Ai}\left[-\left(\frac{2}{n}\right)^{1/3}
\frac{\Omega}{\omega}\right]-{\rm i}({\bar s}_\varphi
{\hat{\bf e}}_\parallel-{\bar s}_r\cos\theta_P
{\hat{\bf e}}_\perp)\nonumber\\
&&\times\left(\frac{2}{n}\right)^{1/3}{\rm Ai}^\prime
\left[-\left(\frac{2}{n}\right)^{1/3}\frac{\Omega}{\omega}\right],
\label{eq:a24}
\end{eqnarray}
and
\begin{equation}
{\bar s}_{r,\varphi,z}\equiv\int_{-\infty}^\infty {\rm d}{\hat z}\,
\exp({\rm i}n{\hat z}\cos\theta_P)s_{r,\varphi,z}\big
\vert_{{\hat r}=\csc\theta_P}.
\label{eq:a25}
\end{equation}
Here, $\Re$ stands for the real part of the quantity inside brackets, 
$s_{r,\varphi,z}$ are the cylindrical components of ${\bf s}$, 
and ${\hat r}_<$ and ${\hat r}_>$ denote the radial boundaries 
of the superluminal part of the source that contributes towards the 
radiation at $P$~\citep{b10}. 

As in the case of the ring source, the fact that the radiation 
observed at the time $t_P$ is determined almost exclusively 
by the state of the emitting current at the retarded time 
$t=t_c$ turns the temporal modulation $\exp({\rm i}\Omega t)$ 
of the source into a spatial modulation 
$\exp(-{\rm i}\Omega\varphi_0/\omega)$.  
Consequently, the remaining integral in equation 
(\ref{eq:a21}) again has a non-zero value for all $n$ 
when $\Omega/\omega$ differs from an integer. 
Insertion of the resulting expression for ${\tilde{\bf E}}_n$ 
in ${\rm d}P_n/d\Omega_P=c{R_P}^2\vert{\tilde{\bf E}}_n\vert^2/(2\pi)$ 
now yields the radiated power per harmonic per unit solid angle:
\begin{eqnarray}
\frac{{\rm d}P_n}{{\rm d}\Omega_n}&\simeq&\frac{c^3}{2\pi\omega^2}
\csc^2\theta_P\vert K_r\vert^2 {K_{\varphi_0}}^2
\left(\frac{n}{2}\right)^{4/3}\vert{\bf V}\vert^2,
\label{eq:a26}
\end{eqnarray}
where
\begin{equation}
K_{\varphi_0}=(-1)^{n+m}\sin\left(\frac{\pi\Omega}{\omega}\right)
\left(\frac{\mu_+}{n-\mu_+}+\frac{\mu_-}{n-\mu_-}\right),
\label{eq:a27}
\end{equation}
and $\mu_\pm=\pm m+\Omega/\omega$.  Since the 
Airy functions in equation (\ref{eq:a24}) decay exponentially 
with $n$ when their argument is positive~\citep{b19}, 
the terms associated with 
$-\Omega<0$ are ignored in the above expression.
\footnote{This expression for ${\rm d}P_n/d\Omega_P$ 
slightly differs from that derived in~\citet{b10} because, here, 
we have eliminated the usual step of integrating the 
first member of equation (\ref{eq:a18}) by parts to transfer
the differentiation with respect to $t$ from the delta function 
onto ${\bf j}$.  Though making no difference in the subluminal 
case, this alternative procedure has the effect, in the 
superluminal case, of removing a multiplicative factor 
$\mu_\pm$ from $K_{\varphi_0}$ and instead 
multiplying the expression for ${\tilde{\bf E}}_n$ by $n$.}

Evaluating the absolute value of the complex vector ${\bf V}$ in
equation (\ref{eq:a24}), we obtain
\begin{eqnarray}
\frac{{\rm d}P_n}{{\rm d}\Omega_P}& \propto &{S_1(n)}^2
{\rm Ai}^2\left[-\left(\frac{2}{n}\right)^{1/3}
\frac{\Omega}{\omega}\right]\nonumber\\
&&\mbox{}+{S_2(n)}^2\left(\frac{2}{n}\right)^{2/3}
{\rm Ai^\prime}^2\left[-\left(\frac{2}{n}\right)^{1/3}
\frac{\Omega}{\omega}\right]\nonumber\\
&&\mbox{}+2{S_3(n)}^2\left(\frac{2}{n}\right)^{1/3}{\rm Ai}
\left[-\left(\frac{2}{n}\right)^{1/3}
\frac{\Omega}{\omega}\right]\nonumber\\
&&\mbox{}\times{\rm Ai^\prime}\left[-\left(\frac{2}{n}
\right)^{1/3}\frac{\Omega}{\omega}\right],
\label{eq:a28}
\end{eqnarray}
in which
\begin{equation}
S_1(n)=n^{2/3}\vert K_r\vert K_{\varphi_0}\left(\vert{\bar s}_r
\vert^2+\vert{\bar s}_\varphi\cos\theta_P-{\bar s}_z\sin\theta_P
\vert^2\right)^{1/2},
\label{eq:a29}
\end{equation}
\begin{equation}
S_2(n)=n^{2/3}\vert K_r\vert K_{\varphi_0}\left(\vert{\bar s}_\varphi
\vert^2+\vert{\bar s}_r\vert^2\cos^2\theta_P\right)^{1/2},
\label{eq:a30}
\end{equation}
and
\begin{eqnarray}
S_3(n)& = &n^{2/3}\vert K_r\vert K_{\varphi_0}\left
\{\Im\left[{{\bar s}_r}^*\cos\theta_P\left({\bar s}_\varphi
\cos\theta_P\right.\right.\right. \nonumber\\
&&\left.\left.\left. -{\bar s}_z\sin\theta_P\right)-{\bar s}_r
{{\bar s}_\varphi}^*\right]\right\}^{1/2}.
\label{eq:a31}
\end{eqnarray}
Here, $\Im$ and the superscript star denote the imaginary 
part and the conjugate of a complex variable, respectively.

\label{lastpage}

\end{document}